\documentclass{aa}

\usepackage{natbib}
\bibpunct{(}{)}{;}{a}{}{,} 
\usepackage{graphics}   
\usepackage{graphicx}   
\usepackage{epstopdf}
\usepackage{longtable}   
\usepackage{url}      
\usepackage{bm}        
\usepackage[varg]{txfonts}
\usepackage{pdflscape}
\usepackage{supertabular}
\usepackage{epsfig,psfrag,epic,eepic}
\usepackage[bookmarks=false,pdftoolbar=true,pdfmenubar=true,colorlinks=true,linkcolor=blue,citecolor=blue]{hyperref}   

\begin{document}

\title{Star-planet interactions} 
\subtitle{I. Stellar rotation and planetary orbits}

\author{Giovanni Privitera\inst{1,2}, Georges Meynet\inst{1}, Patrick Eggenberger\inst{1}, Aline A. Vidotto\inst{1,3}, Eva Villaver\inst{4}, 
\and
Michele Bianda\inst{2}
}

 \authorrunning{Privitera et al.}

 \institute{Geneva Observatory, University of Geneva, Maillettes 51, CH-1290 Sauverny, Switzerland
 \and Istituto Ricerche Solari Locarno, Via Patocchi, 6605 Locarno-Monti, Switzerland 
 \and School of Physics, Trinity College Dublin, Dublin-2, Ireland
\and Department of Theoretical Physics, Universidad Autonoma de Madrid, Modulo 8, 28049 Madrid, Spain
}

\date{Accepted}
\abstract {As a star evolves, the planet orbits change with time due to tidal interactions, stellar mass losses, friction and gravitational drag forces, mass accretion and evaporation on/by the planet. 
Stellar rotation modifies the structure of the star and therefore the way these different processes occur. 
Changes of the orbits, at their turn, have an impact on the rotation of the star.} 
{
Models accounting in a consistent way for these interactions between the orbital evolution of the planet and the evolution of the rotation of the star are still missing.
The present work is a first attempt to fill this gap.
}
{We compute the evolution of stellar models including a comprehensive treatment of rotational effects together with the evolution of planetary orbits, so that the exchanges of angular momentum between the star and the planetary orbit are treated in a self-consistent way. 
 The evolution of the rotation of the star accounts for the angular momentum exchange with the planet and also follows the effects of the  internal transport of angular momentum and chemicals.
 These rotating models are computed for initial masses of the host star between 1.5 and 2.5 M$_{\odot}$, with initial surface angular velocities equal to 10 and 50\% the critical velocity on the Zero Age Main Sequence (ZAMS), for a metallicity Z=0.02, with and without tidal interactions with a planet. We consider planets with masses between 1 and 15 Jupiter masses (M$_{J}$) and beginning their evolution at various distances between 0.35 and 4.5 au.
 }
{
We show that rotating stellar models without tidal interactions and without any wind magnetic braking during the red giant phase can well reproduce the surface rotations of the bulk of the red giants.
However, models without any interactions cannot account for fast rotating red giants in the upper part of the red giant branch, where, such models, whatever the initial rotation
considered on the ZAMS, always predict very low velocities.
For those stars some interaction with a companion is highly probable and the present rotating stellar models with planets confirm that tidal interaction can reproduce their high surface velocities.
We show also that the minimum distance between the planet and the star on the ZAMS that will allow the planet to avoid engulfment and survive ({\it i.e.} the survival limit) is decreased around faster rotating stars.
 }
{}
\keywords{Stars: evolution -- Stars: rotation -- Planetary systems}

\maketitle

\titlerunning{Stellar rotation and planet orbital evolution}
\authorrunning{Privitera et al.}

\section{Introduction}

One of the important lessons of the extra-solar planet discoveries is the very large variety of
systems encountered in nature. 
In some cases, tidal forces between the star and the planet could become so large that the semi-major axis of the orbit will decrease leading to the engulfment of the planet by the star. When this occurs, models indicate that changes at the stellar surface could be observed: a transitory and rapid increase in the luminosity \citep{siess99II}, a change in the surface abundance of lithium \citep{carlberg10,adamow12} or an increase of the surface rotation rate \citep{siess99I,siess99II,carlberg09,carlberg10}. 
Interestingly, some stars present observed characteristics that could be related to planet engulfments as
for instance the few percents of red giants (RGs) that are fast rotators \citep{fekel93, massarotti08, carlberg11, carlberg14}.

Many works have studied the physics of star-planet interactions, some of them focusing on how a planetary orbit changes under the action of tides between the star and the planet \citep{livio84, soker84, sackmann93, rasio96, siess99I, siess99II, villaver07, sato08, villaver09, carlberg09, nordhaus10, kunimoto11, bear11, mustill12, nordhaus13, villaver14}. These works have provided many interesting clues about the initial conditions for an engulfment to occur 
in terms of the mass of the star, mass of the planet, initial distance between the planet and the star and of the importance of other physical ingredients as the mass loss rates or the overshooting. One of these works have specifically studied the possibility that these engulfments are the origin of fast rotating red giants \citep{carlberg09}. 
Although the work of \citet{carlberg09} provides a detailed and very interesting discussion of the question, it suffers from a limitation, the fact that its conclusions were based on non-rotating stellar models. This limitation was indeed recognized by the authors\footnote{In their footnote 2, \citet{carlberg09} write: ``Rotation also plays an important role; however, grids of evolution models explicitly accounting for rotational effects are not currently available for the range of stellar masses in our study''.}. The present work is a first attempt to fill this gap by including a comprehensive treatment of rotational effects to compute for the first time the simultaneous evolution of the planetary orbit and of the resulting internal transport of angular momentum and chemicals in the planet-host star. 
In particular, the present approach will provide us with more consistent answers to the following points:

\begin{itemize}
\item What are the surface rotations of red giants predicted by single star models (stars with no interaction with an additional body)? 
A proper comparison between the evolution of the surface velocity with and without tides cannot be made with non-rotating stellar models without a priori assumption on the internal distribution of angular momentum.
Here, this internal distribution is not a priori imposed, but is computed self consistently by following the changes of the structure of the star, the transport of angular momentum by the shear turbulence and the meridional currents, the impact of mass loss and of tides.
\item Rotational mixing changes the size of the cores and, among other features, the time of apparition of the external convective zone that is so important for computing the tidal force according to the formulation by \citet{zahn92}.
It is interesting to see whether these changes are important or not for the computation of the orbital evolution.
\item The evolution of the orbit depends on the ratio between the angular velocity of the outer layers of the star and the orbital angular velocity of the planet (see Eq.~\ref{equa:tides} below); only rotating models can thus account for this effect in a self-consistent manner. 
\end{itemize}
In the present paper, we focus on the effects of stellar rotation on the evolution of the planetary orbit and on the impact of the changes of the planetary orbit on the rotation of the star. In a second paper, we will discuss the evolution of the rotating star after the engulfment.
In Sect. \ref{sec:2}, we describe the physics included in our models for the stars, the orbits and the planets, emphasizing on the new aspects of our approach with respect to previous works. The evolution of rotating models evolved without interaction with a planet are described in Sect. \ref{sect3}. The exchanges of angular momentum between the star and the planetary orbit are discussed in Sect. \ref{sec:4}, while the main points learned from this study are summarized in Sect. \ref{sect5}.

\section{Ingredients of the models}\label{sec:2}

\subsection{Stellar models} 
Stellar models are computed with the Geneva stellar evolution code, which includes a comprehensive treatment of rotational effects \citep[see][for a detailed description]{egg08}. The ingredients of these models (nuclear reaction rates, opacities, mass loss rates, initial composition, overshooting, diffusion coefficients for rotation) are computed as in \citet{ekstrom12}. The reader can refer to this reference for all the details; we just recall here some of the main points:
\begin{itemize}
\item \emph{Convection} and \emph{overshooting}: convective regions are determined using the Schwarzschild criterion. An overshoot parameter $d_{over}/H_{p}$ is used to extend the convective core \citep[see][]{ekstrom12}, with a value $d_{over}/H_{p}=0.05$ for $1.5\ M_{\odot}$ stars and $d_{over}/H_{p}=0.1$ for more massive stars. The outer convective zone is treated according to the mixing length theory, with a value for the mixing-length parameter, $\alpha=l/H_p$,  equal to 1.6.
\item \emph{Mass loss rate}: 
the mass loss is non negligible during the RG branch. We used the prescription by \citet{rei75}
\begin{equation}
\dot{M}_{loss}= 4\times10^{-13}\eta L_{\star}R_{\star}/M_{\star} \ \ M_{\odot}\ {\rm yr}^{-1}\ \ \ \ ,
\label{equa:mass_loss_rate}
\end{equation}
with $\eta = 0.5$ \citep{maeder89}. $L_{\star}$, $R_{\star}$ and $M_{\star}$ are the luminosity, the radius and the mass of the star. The mass loss is one of the parameters that drives the planetary orbital evolution.
\item \emph{Shear} and \emph{meridional currents}: rotating models are computed using the assumption of shellular rotation \citep{zahn92}, which postulates that the angular velocity remains nearly constant along an isobar in differentially rotating stars, due to a strong horizontal turbulence. The prescription of \citet{zahn92} is used for this strong horizontal diffusion coefficient. The expressions for the meridional velocity, the coefficients corresponding to the vertical shear and to the transport of chemicals through the combined action of meridional currents and horizontal turbulence are taken as in \citet{ekstrom12}. 
\end{itemize}

The transport of angular momentum inside a star is implemented following the prescription of  \citet{zahn92}.
This prescription was complemented by \citet{MZ1998}. In the radial direction, it obeys the equation
\begin{equation}
\label{transmom}
\rho \frac{d}{d t}( r^{2} \bar{\Omega})_{M_r} =
   \frac{1}{5r^2} \frac{\partial}{\partial r} ( \rho r^4 \bar{\Omega} U(r) )
   + \frac{1}{r^2} \frac{\partial}{\partial r} ( \rho D r^4 \frac{\partial \bar{\Omega}}{\partial r}) \, ,
\end{equation}
where $\bar{\Omega}$ is the mean angular velocity on an isobaric surface, $r$ the radius, $\rho$ the density, $M_r$ the mass inside the radius $r$, $U$ the amplitude of the radial component of the meridional circulation\footnote{The radial component
$u(r,\theta)$ of the velocity of the meridional circulation at a distance $r$ to the center and at a colatitude $\theta$ can be written $u(r,\theta)=U(r)P_2(\cos \theta)$, where $P_2(\cos \theta)$ is the second Legendre polynomial. Only the radial term $U(r)$ appears in Eq.~2.}, and $D$ the total diffusion coefficient in the vertical direction taking into account the various instabilities that transport angular momentum.
The first term on the right hand side of this equation is the divergence of the advected flux of angular momentum, while the second term is the divergence of the diffused flux. The effects of expansion or contraction are automatically
included  in a Lagrangian treatment. The expression of $U(r)$ \citep[see][]{MZ1998} involves
derivatives up to the third order; Eq. \ref{transmom} is thus of the fourth order and implies
four boundary conditions. These conditions are obtained by requiring momentum conservation and the absence of differential rotation at
convective boundaries \citep{tal97}. In particular, the boundary condition imposing momentum conservation at the bottom of the convective envelope has to take into account the impact of tides due to the presence of the planet: 
\begin{equation}
{\partial \bar{\Omega} \over \partial t}
 \int _{R_{\rm env}}^{R_{\star}}  r^4 \rho \, {\rm d} r = -\frac{1}{5} R_{\rm env}^4 \rho \bar{\Omega} U + {\cal F}_\Omega   \, ,
\end{equation}
with $R_{\rm env}$ the radius at the base of the convective envelope. ${\cal F}_\Omega$ represents the torque
applied at the surface of the star. It is given here by:
\begin{equation}
{\cal F}_\Omega={{\rm d} (\Omega_{\star} I_{\rm ce})\over{\rm d}t} =-{1 \over 2}  {\cal L}_{\rm pl} \left( {\dot{a} \over a}\right)_{t} \, ,
\end{equation}
where $\Omega_{\star}$ is the angular velocity at the surface of the star and $I_{\rm ce}$ is the moment of inertia of the convective envelope,
${\cal L}_{\rm pl}$ is the angular momentum of the planetary orbit and $(\dot{a}/a)_{t}$ is the inverse of the timescale for the change of the orbit
of the planet resulting from tidal interaction between the star and the planet. 
The expression of $(\dot{a}/a)_{t}$ is discussed below.

\subsection{Physics of the evolution of the orbit}
The evolution of the semi-major axis $a$ of the planetary orbit, that we suppose circular ($e=0$) and aligned with the equator of the star, is given by \citep[see][]{zahn66,alexander76,zahn77,zahn89,livio_soker84,villaver09,mustill12,villaver14}
\begin{equation}
\left(\frac{\dot{a}}{a} \right)=
\underbrace{-\frac{\dot{M}_{\star}+\dot{M}_{pl}}{M_{\star}+M_{pl}}}_{\rm Term\  1}
\underbrace{-\frac{2}{M_{pl}v_{pl}}\left[F_{fri} + F_{gra}\right]}_{\rm Term\  2}
\underbrace{-\left(\frac{\dot{a}}{a} \right)_{t}}_{\rm Term\  3}
\ \ \ \ ,
\label{equa:evoorb}
\end{equation}
where $\dot{M}_{\star}=-\dot{M}_{loss}$ with $\dot{M}_{loss}$ being the mass loss rate (here given as a positive quantity). $M_{pl}$ and $\dot{M}_{pl}$ are the planetary mass and the rate of change in the planetary mass, $v_{pl}$ is the velocity of the planet. $F_{fri}$ and $F_{gra}$ are respectively the frictional and gravitational drag forces, while $(\dot{a}/a)_{t}$ is the term that takes into account the effects of the tidal forces. 
We shall not discuss here the expressions of terms 1 and 2, since they are already extensively discussed in the references mentioned above.
Here we focus on term 3, the tidal term, which is the term responsible for the exchange of angular momentum between the planet and the star and in which the rotation of the star is explicitly involved. 

When a convective envelope appears, tidal dissipation can be very efficient in the stellar envelope. As a result, angular momentum is transferred from the planetary orbit to the star or the inverse depending on whether the orbital angular velocity of the planet is smaller or larger than the axial angular rotation of the star.
The term 3 is given by \citep{zahn66,zahn77,zahn89}:
\begin{equation}
(\dot{a}/a)_{t}=\frac{f}{\tau}\frac{M_{env}}{M_{\star}}q(1+q) \left(\frac{R_{\star}}{a}\right)^{8}\left(\frac{\Omega_{\star}}{\omega_{pl}}-1\right)\ \ \ \ ,
\label{equa:tides}
\end{equation}
with $M_{env}$ the mass of the convective envelope, $q=M_{pl}/M_{\star}$, $\omega_{pl}$ the orbital angular velocity of the planet, and $\Omega_{\star}$ the angular velocity at the stellar surface. $\tau$ is the convective eddy turnover timescale and $f$ is a numerical factor equal to the ratio of the orbit half period $P/2$ to the convective eddy turnover time $f=(P/2\tau)^2$ when $\tau>P/2$ in order to consider the only convective cells that give a contribution to the viscosity; otherwise $f$ is equal to 1 \citep{villaver09}. The eddy turnover timescale is taken as in \cite{rasio96}:
\begin{equation}
\tau=\left[\frac{M_{\rm env}(R-R_{\rm env})^{2}}{3L_{\star}} \right]^{1/3}\ \ \ \ .
\label{equa:tau}
\end{equation}

\subsection{The planet model}\label{sec:pl_phen}

The frictional drag force depends on the radius of the planet. 
We assume that the planet/brown dwarf is a polytropic gaseous sphere of index $n=1.5$ \citep{siess99I}. We use the mass-radius relation by \citet{zapolsky69}.
In our computations, we have taken into account for the first time the fact that the effective radius of the planet may be higher due to a planetary magnetic field \citep{Vidotto2014}. We use the magnetic pressure radius given by 
\begin{equation}
R_{mp}=R_{pl}\left(\frac{B_{pl}^{2}}{8\pi\rho v_{wind}^{2}}\right)^{1/6}\ \ \ \ ,
\label{equa:Rmp}
\end{equation}
where $B_{pl}$ is the dipole's magnetic field strength of the planet \citep{chapman30,cambridge11}.
For $B_{pl}$ we use the magnetic field of Jupiter at the equator that is equal to 4.28 Gauss.

In section A.2.1 of Appendix A, we show that the magnetic radius can be about a hundred times larger than the radius of the planet. 
This increases the frictional term by about 4 orders of magnitudes.
However, even in that case, friction can only change the radius of the orbit by about one percent. Thus, the impact of the friction term remains small.

\subsection{Initial conditions considered}

We consider stars with initial masses of 1.5, 1.7, 2 and 2.5 M$ _{\odot}$ and an initial rotation equal to $\Omega_{ini}/\Omega_{crit}=0.1$ and 0.5, where $\Omega_{ini}$ is the initial angular velocity on the ZAMS and $\Omega_{crit}$ the critical angular velocity on the ZAMS ({\it i.e.} the angular velocity such that the centrifugal acceleration at the equator balances the acceleration due to the gravity at the equator). 

The initial mass range considered in this work (M $>1.5$ M$_\odot$) contains relatively fast rotators (see next section) in contrast with lower initial mass stars (M $<1.5$ M$_\odot$). Indeed, stars above 1.5 M$_{\odot}$ do not have a sufficiently extended outer convective zone to activate a dynamo during the main sequence, so that unless they host a fossile magnetic field, they do not undergo any significant magnetic braking. Presently only a small fraction of these stars \citep[of the order of 5-10\%, see the review by][and references therein]{Donati2009} host a surface magnetic field between 300 G and 30 kG. Lower initial mass stars have an extended outer convective zone during the main sequence and thus activate a dynamo and suffer a strong braking of the surface by magnetized winds. The physics of these lower initial mass stars is therefore different and will be the topic of another paper in this series.

A metallicity of $Z=0.02$ has been chosen to account for the fact that the mean metallicity of the current sample of planet-host stars is slightly higher than solar \citep{santos2001,santos2004,sousa2011} and so it is the metallicity of the RG stars known to host planets in the mass range under study \citep{maldonado13, Maldonado2016}.
Planets with masses equal to 1, 5, 10 and 15 Jupiter masses ($M_{J}$) have been considered.
The initial semi-major axes ($a_{0}$) have been taken in the range [0.35-4.5] au. 
The eccentricities of the orbits are fixed to 0. The computations were performed until the He-flash (tip of the RG branch) for the 1.5, 1.7 and 2 M$_\odot$, 
and until the end of the core He-burning phase for the 2.5 M$_{\odot}$\footnote{This model ignites helium in a non-degenerate regime, there is therefore no helium flash.}.

\section{Rotating stellar models without tidal interactions}\label{sect3}

It is important to study the evolution of single rotating stars in order to reveal, by contrast, the differences that appear when tidal interactions with a planet are occurring. In this section, we first describe how the surface velocity of single star evolves
from the Main-Sequence (MS) phase up to the tip of the RG branch. We then discuss how the corotation radius evolves, because this quantity plays a key role to determine in which direction the angular momentum is transferred, from the orbit to the star or from the star to the orbit. Finally, we study how rotation, by changing the global and internal properties of the star, has an impact on the tidal forces and hence on the evolution of a planetary orbit and of its own rotation. 

\subsection{Evolution of the surface rotation for isolated stars}

\begin{figure*}
\centering
\includegraphics[width=.49\textwidth, angle=0]{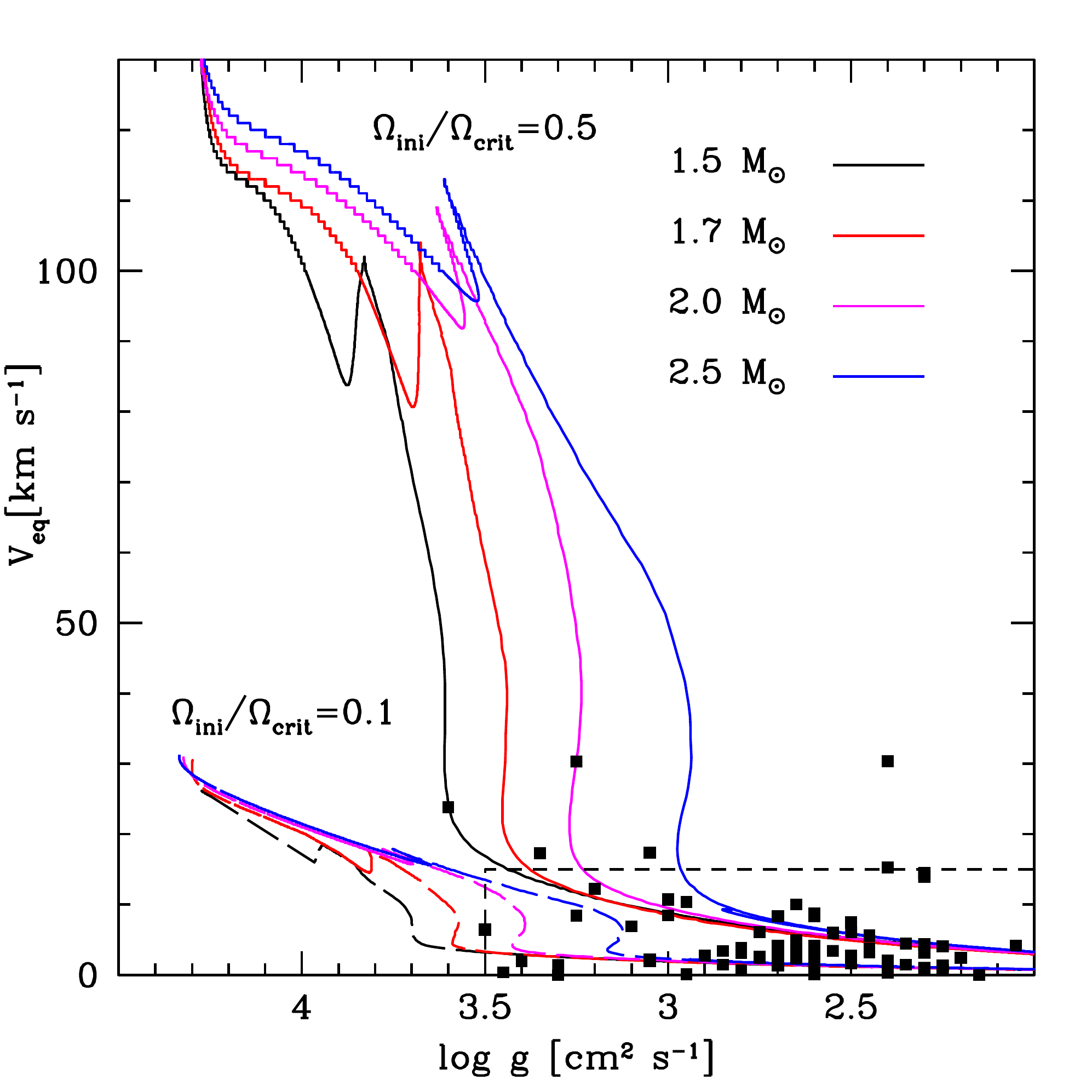}\includegraphics[width=.49\textwidth, angle=0]{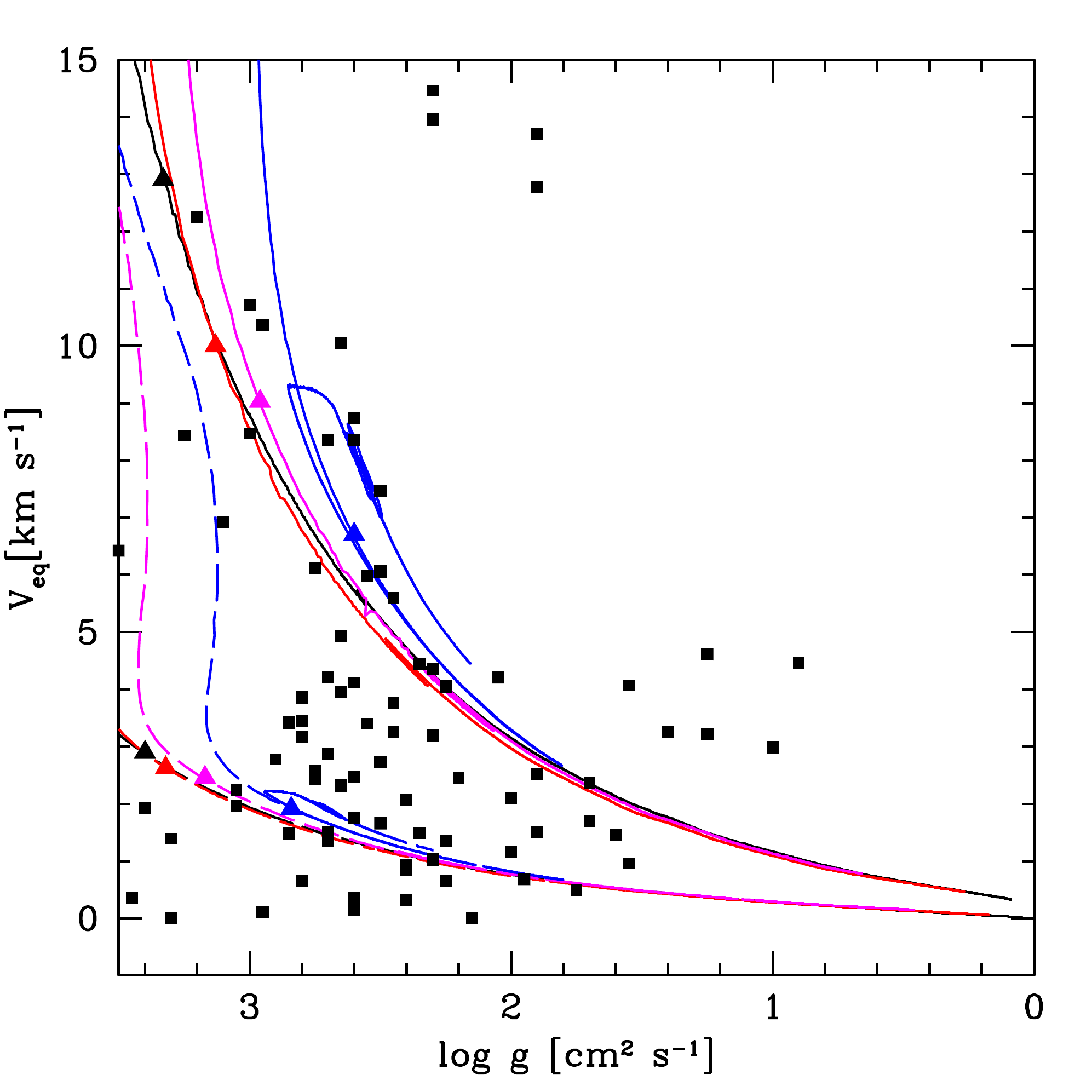}
\includegraphics[width=.49\textwidth, angle=0]{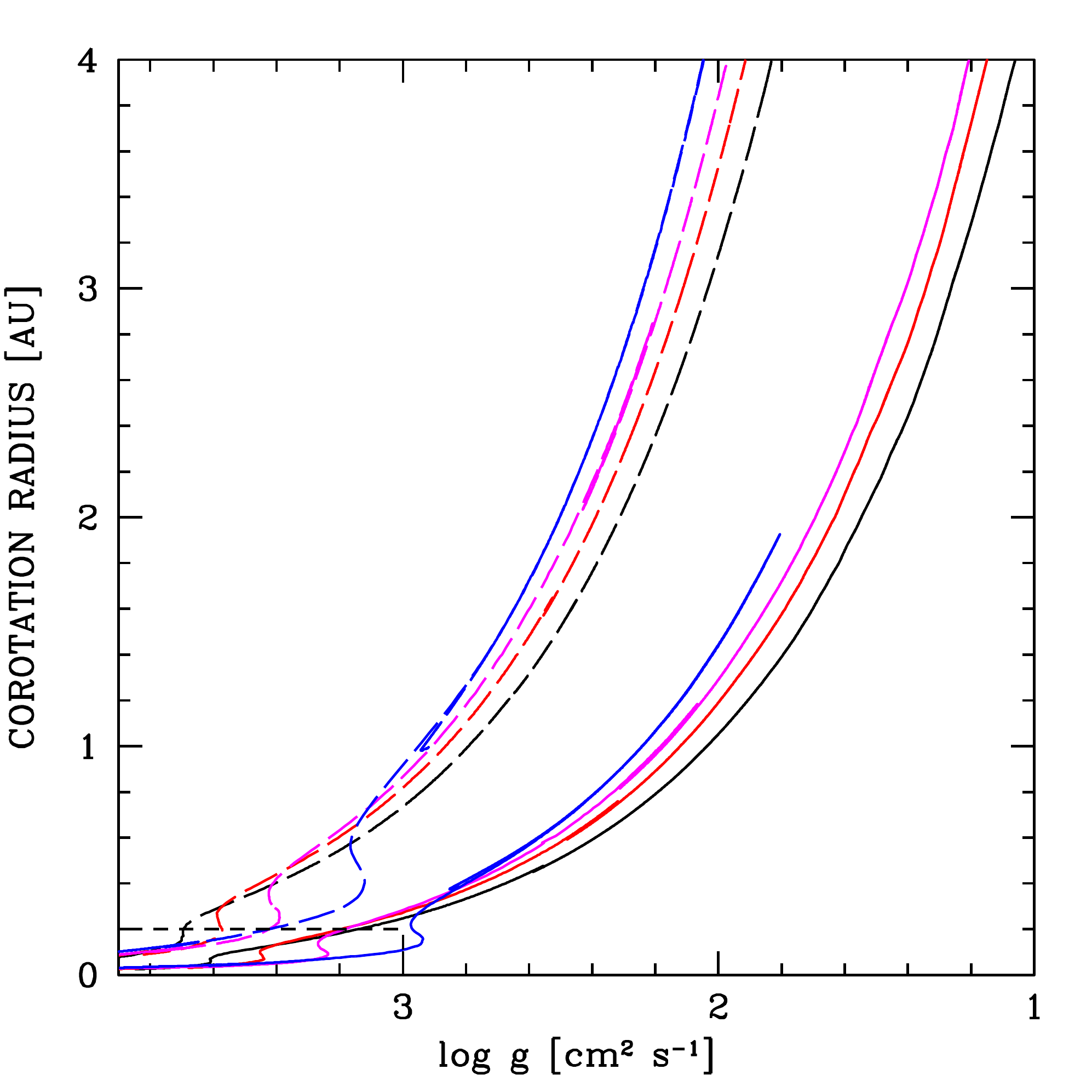}\includegraphics[width=.49\textwidth, angle=0]{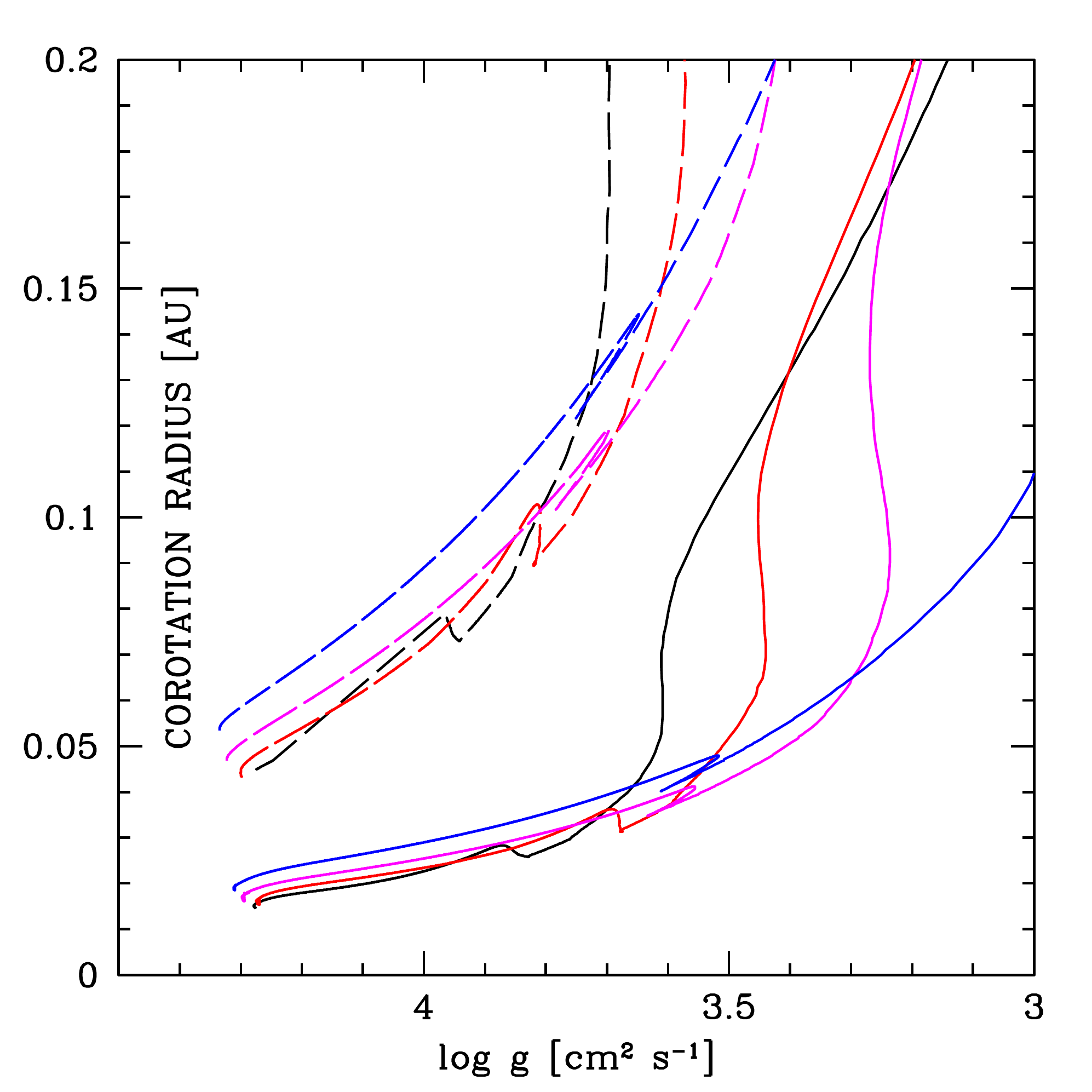}

\caption{
{\it Upper left panel:} Evolution of the surface equatorial velocity for our stellar models. No tidal interactions and planet engulfments are considered. The initial masses and rotations are indicated. The rectangle area indicated at the right bottom corner indicates the region
that is zoomed on the upper right panel.
{\it Upper right panel:} Zoom on the RG branch phase for the same stellar models as those shown on the upper left panel. Triangles indicate where the dredge-up occurs. Black full squares correspond to the sample of red giants observed by \protect\citet{carlberg12}.
{\it Lower left panel:} Evolution of the corotation radius (in au) for rotating stellar models with no tidal interactions during the RG branch phase. For the 2.5 M$_\odot$, the core He-burning phase is also shown. The rectangle area indicated at the left bottom corner indicates the region
that is zoomed on the lower right panel.
{\it Lower right panel:} Same as the lower left panel, but the MS phase and the crossing of the Hertzsprung-Russel (HR) gap are shown.
}
\label{fig:vit}
\end{figure*}

The upper left panel of Fig.~\ref{fig:vit} shows the evolution of the equatorial surface velocity from the ZAMS until the tip of the RG branch or the end of the core He-burning phase (for the 2.5 M$_\odot$) for stellar models without any tidal interactions.  We see that our slow rotating models ($\Omega_{\rm ini}/\Omega_{\rm crit}$=0.1) correspond to surface rotations during the MS phase around 20 km s$^{-1}$, while
our fast rotating models ($\Omega_{\rm ini}/\Omega_{\rm crit}$=0.5) correspond to surface velocities around 100-110 km s$^{-1}$. 

\citet{ZR2012} analysed the $\upsilon\sin i$ for a sample of more than two thousand B6- to F-type stars. They 
find that for stars with masses between about 1.6-2.5 M$_\odot$, the velocity distribution is unimodal, with maxwellian distributions that peak between 160 and 180 km s$^{-1}$. The adopted rotation rates are thus in the low range of observed velocities. We focus on this low range of rotations for mainly the following reason: stars with planets might show slower initial rotations than stars without planets, since the angular momentum of the protostellar cloud has to be shared between the planets and the star instead of being locked in the star only. Of course, many processes may evacuate the angular momentum of the cloud, planet formation being only one of them. Nevertheless, it appears reasonable to think that stars with planets might initially rotate slower than stars without planets \citep{Bouvier2008}. 

The upper right panel of Fig.~\ref{fig:vit} shows that along the RG branch the surface velocity drops to lower and lower values when the surface gravity decreases. The bulk of red giants observed by \citet{carlberg12} are characterized by initial masses (estimated from their positions in the HR diagram) between 1.3 and 3 M$_\odot$. They are well framed by our slow and fast rotating models despite the fact that, as discussed just above, our initial rotations span only a subset of the range of values shown by the progenitors of red giants. This mainly comes from the fact that the inflation of the star during the RG phase is so large that many different initial surface rotations on the ZAMS converge to similar values at that stage. We see also that to explain the slow observed rotation of the bulk of the red giant stars, there is a priori no need to invoke any magnetic braking that would be activated when the convective envelope appears. 

The big triangles along the tracks indicate where the dredge-up occurs. No surface acceleration is observed at that point. Therefore, our models do not confirm the idea suggested by \citet{simon89} that a short-lived rapid rotation phase during the RG phase occurs when the deepening stellar convection layer dredges up angular momentum from the more rapidly rotating stellar interior. Actually, some angular momentum is indeed dredged-up from the core to the surface, but the central region of the star is so compact and its moment of inertia so small that, even if the core rotates fast, its angular momentum is quite small with respect to the angular momentum of the envelope. Thus, this dredge-up has a negligible impact on the surface rotation of red giants.

Finally, let us note that among the 17 red giants with a $\upsilon\sin i$ above 8 km s$^{-1}$, about 7 (those at the base of the RG branch in the vicinity of the $\Omega_{\rm ini}/\Omega_{\rm crit}$=0.5 tracks) might be explained without any particular acceleration mechanism. Of course, $\upsilon\sin i$ is a lower limit to the true equatorial surface velocity and thus among these stars a few stars can still be much more rapid rotators, but without any other pieces of information, these surface velocities cannot be considered as strong evidences of tidal interaction with a planet or a brown dwarf, or resulting from an engulfment. Stronger candidates are those red giants that are well above the tracks.
These best candidates are not located at the base of the RG branch. This differs from conclusions obtained in previous works, where it was suggested that the rapid rotation signal from ingested planets is most likely to be seen on the lower RG branch \citep{carlberg09}. 
Also the present results show that the use of a fixed lower limit for $\upsilon\sin i$ around 8 km s$^{-1}$ does not appear very adequate to characterize red giants that are candidates for tidal interaction with a planet or a brown dwarf, or resulting from an engulfment. This limit clearly depends on the surface gravity for a given initial mass star.

\subsection{Evolution of the corotation radius}

The corotation radius corresponds to the orbital radius for which the orbital period would be equal to the rotation period of the star. When the actual distance of the planet to the star is inferior to the corotation radius, tidal forces reduce the orbital radius, while when the actual distance of the planet to the star is larger, tidal forces enlarge the orbital radius. This is accounted for in Eq.~\ref{equa:tides} through the term $\left(\frac{\Omega_{\star}}{\omega_{pl}}-1\right)$.

The two lower panels of Fig.~\ref{fig:vit} show the evolution of the corotation radius ($D_{\rm corot}=(G M/\Omega_*^2)^{1/3}$)
for various rotating models (without planets). The lower right panel shows the situation during the MS phase and the crossing of the HR gap, while the lower left panel shows the evolution along the RG branch (and in case of the 2.5 M$_\odot$ also during the core He-burning phase).

In order to have an engulfment, a necessary condition (but of course not a sufficient one) is that the actual radius of the planetary orbit is inferior to the corotation radius. For an engulfment to occur, it is in addition needed that the tidal forces, when the orbital radius is smaller than the corotation radius, are of sufficient amplitude.

The corotation radius is very small during the MS phase. Even in case tidal dissipation would be efficient at that stage, tidal forces can only decrease the radius of the orbit when the distance is below 0.05-0.15 au if the star has a quite low initial rotation rate ($\Omega_{\rm ini}/\Omega_{\rm crit}=0.1$). 
If the star is initially rotating rapidly ($\Omega_{\rm ini}/\Omega_{\rm crit}>0.5$),
in case of the 2 M$_\odot$, the corotation radius is inferior to 0.01 and 0.03 au during the MS phase\footnote{As discussed above, we do not expect frequent magnetic braking during the MS phase since these stars, having no extended outer convective envelope do not present any
efficient dynamo activity.}.

Along the RG branch, the corotation radius increases a lot as a result of the decrease of the surface rotation rate. The corotation radii are shifted downwards when the initial rotation rate increases, as was the case during the MS phase. 

The initial distances between the planet and the star considered in this work are clearly above the corotation radius during the MS phase and will cross it
and thus become inferior to it during the red giant branch. As we shall see, since the rotation of the star is modified due to the tidal interaction, we shall have to see
how the corotation radius will change as a result of this interaction (see Sect.~4.3).

\subsection{Impact of rotation on the structure and the evolution of the star}

\begin{figure}
\includegraphics[width=.49\textwidth, angle=0]{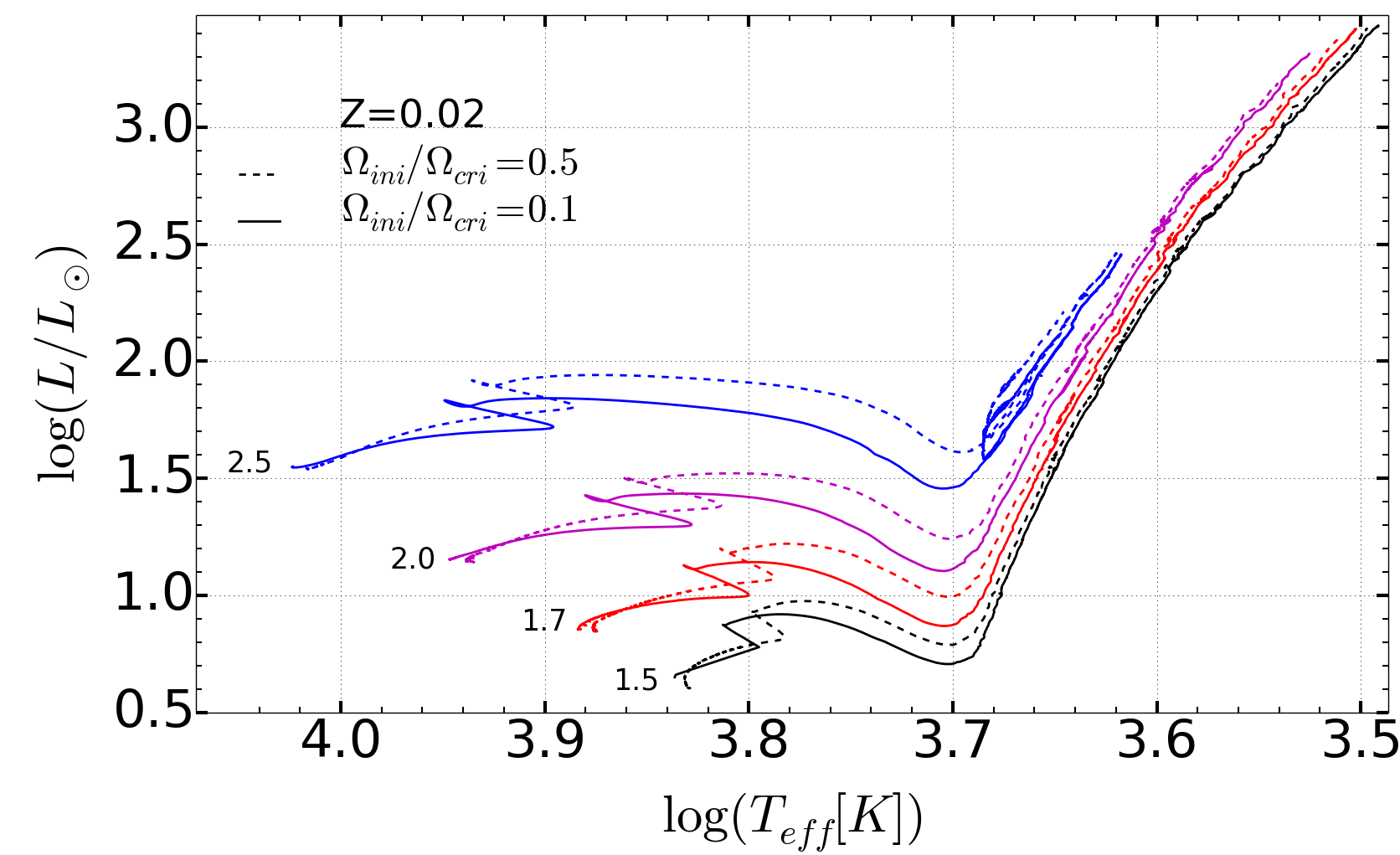}
\caption{
Evolutionary tracks in the Hertzsprung-Russell diagram for
rotating models of 1.5, 1.7, 2.0 and 2.5 M$_{\odot}$. The solid and dashed lines indicate models with $\Omega_{ini}/\Omega_{crit}=0.1$ and $0.5$, respectively. 
} 
\label{fig:HR_diagram}
\end{figure}

\begin{figure}
\centering
\includegraphics[width=.49\textwidth, angle=0]{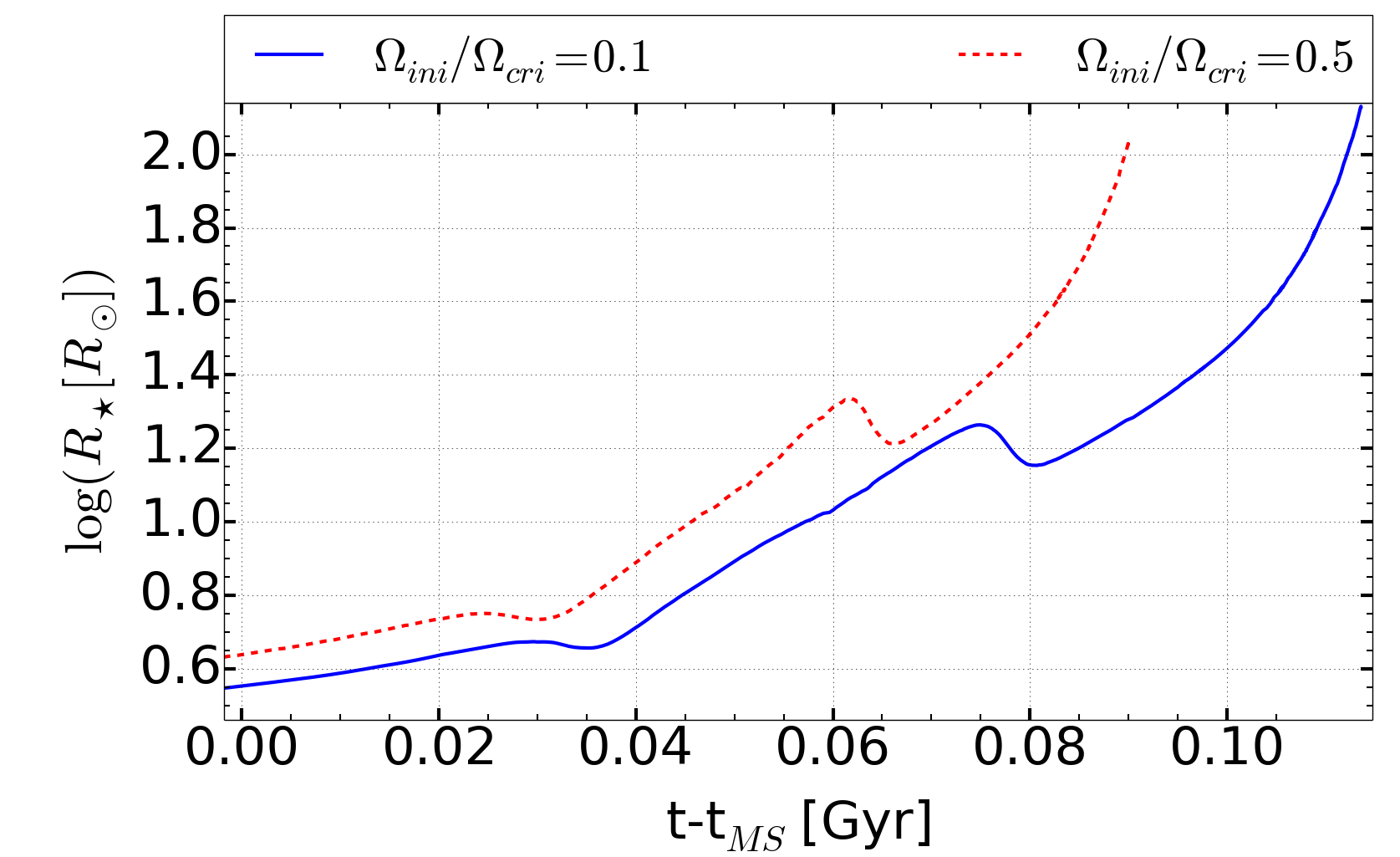}

\includegraphics[width=.49\textwidth, angle=0]{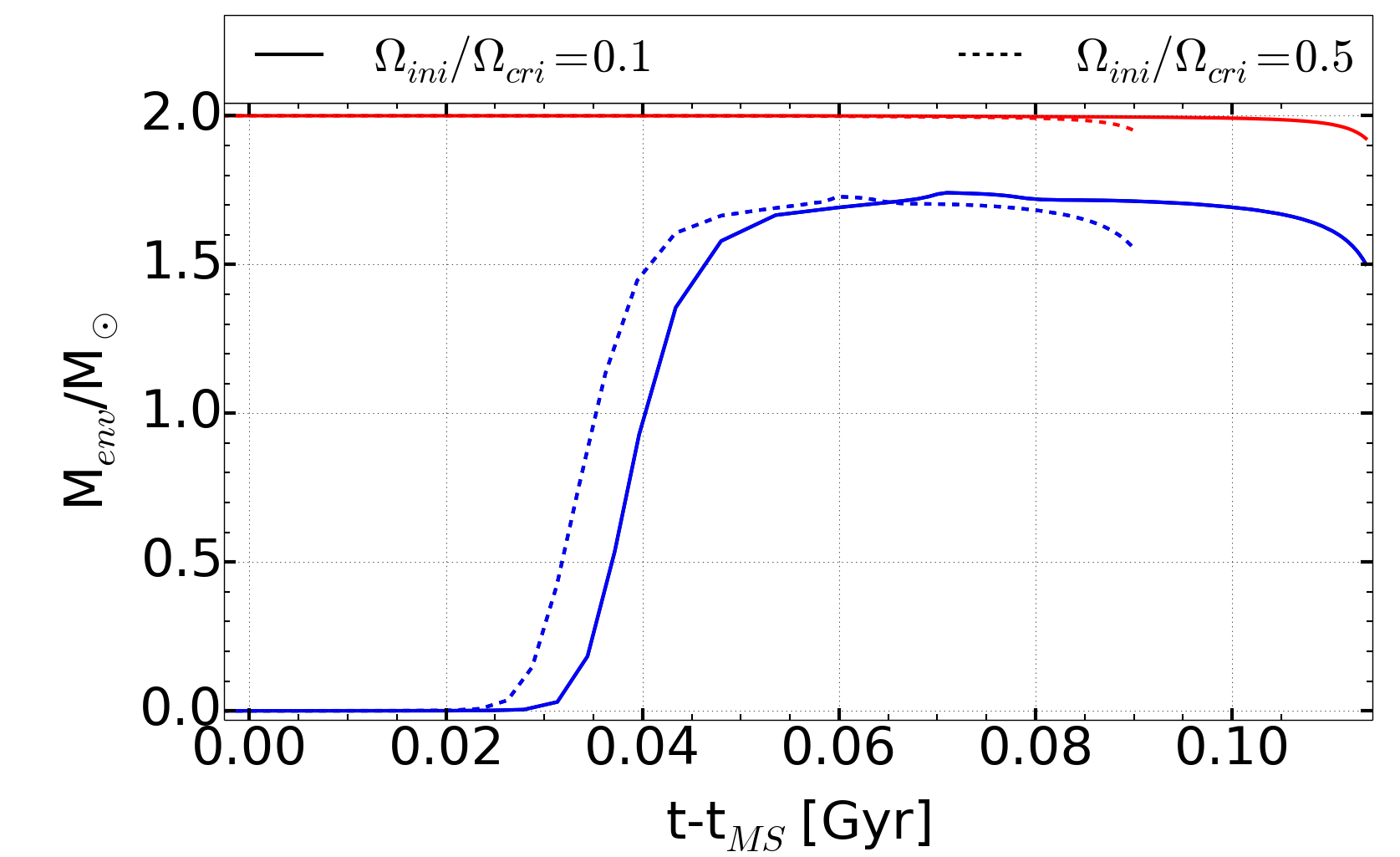}

\includegraphics[width=.49\textwidth, angle=0]{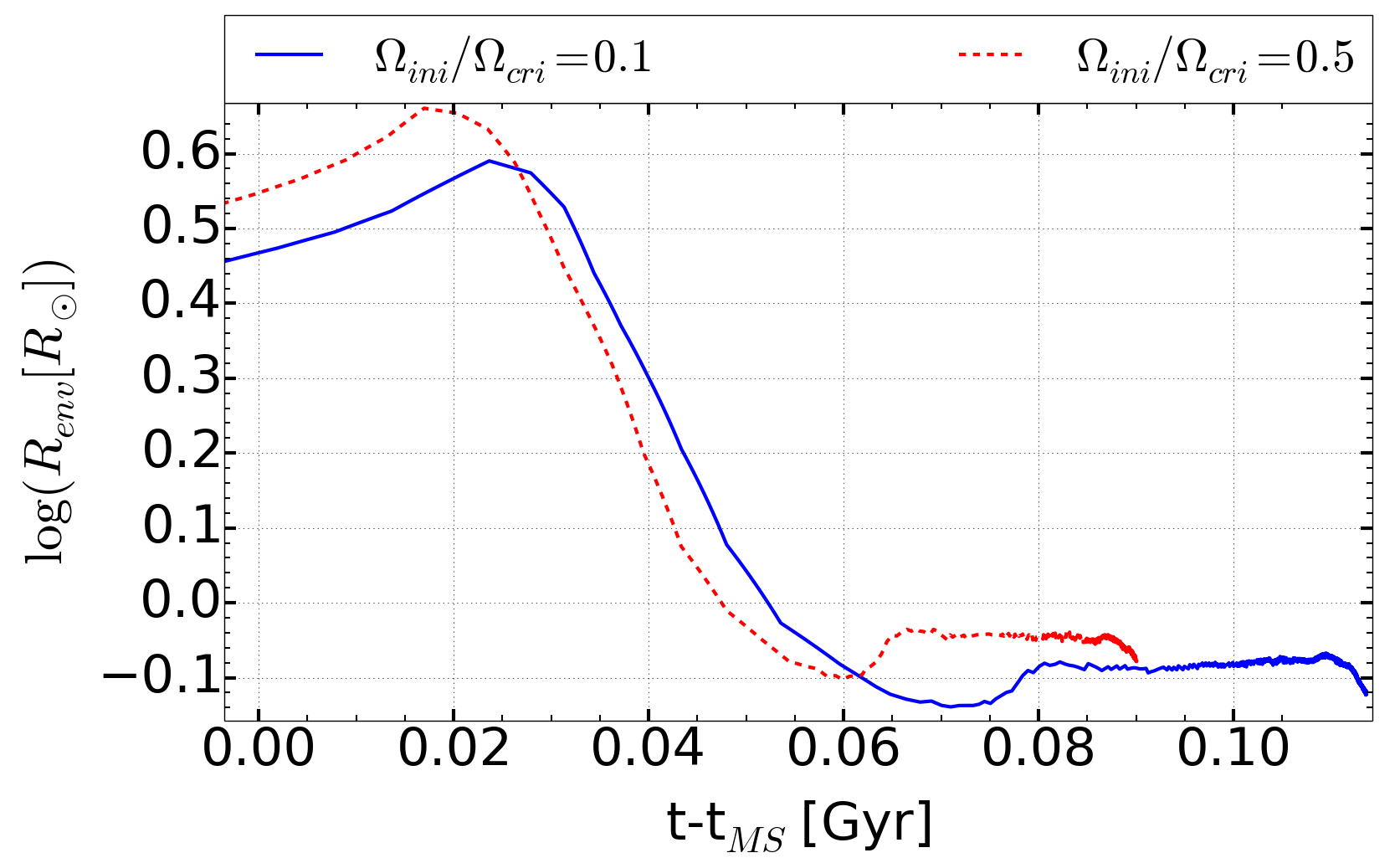}
\caption{
{\it Top:} Evolution of stellar radii for the 2 M$_{\odot}$ models. Continuous and dotted lines correspond to stars with initial velocities $\Omega_{ini}/\Omega_{crit}=0.1$ and $0.5$, respectively. Time 0 corresponds to the end of the MS phase.
{\it Center:} Evolution of the total mass of the star (upper red lines) and of the masses of the convective envelopes (lower blue curves) for the same stars as in the left panel.
{\it Bottomt:} Radii at the bottom of the convective envelopes for the same stars as in the left panel.}
\label{fig:struc}
\end{figure}

In Fig. \ref{fig:HR_diagram}, the evolutionary tracks in the HR diagram of our stellar models are shown. Differences between rotating models for a given mass appear mainly during the MS phase and during the crossing of the HR gap. Faster rotating models are overluminous at a given mass and the MS phase extends to lower effective temperatures. The widening of the MS is mainly due to the increase of the convective core related to the transport by rotational mixing of fresh hydrogen fuel in the central layers. The impact on the luminosity results from both the increase of the convective core and the transport of helium and other H-burning products into the radiative zone \citep[e.g.][]{eggen10,MM2012}. The increase of the convective core also leads to an increase of the MS lifetime. Typically, faster rotating models will reach a given luminosity along the RG branch at an older age than the slower rotating models.

Along the RG branch, at a given effective temperature, the initially fast rotating track is slightly overluminous with respect to the slow one. This means that the radius of the fast rotating star is also larger than the radius of the slow one. The tidal forces
will therefore be stronger around the fast rotating star (the tidal force varies with the stellar radius at the power 8) and we may expect that everything being equal, when an engulfment occurs, it will occur at an earlier evolutionary stage (typically at smaller luminosities along the RGB) around fast rotating stars than around slow rotating ones. Note that since rotation increases the MS lifetime, a given evolutionary stage along the RGB occurs at a larger age when rotation increases.
We also note that the tip of the RG branch occurs at slightly lower luminosities for the faster rotating models (see Fig.~\ref{fig:HR_diagram}). This implies that the maximum radius reached by the star decreases when the initial rotation of the star increases. Therefore, increasing the stellar rotation lowers a bit the maximum initial distance between the star and the planet that leads to an engulfment\footnote{Indeed, the larger the maximum stellar radius, the stronger the tidal forces and thus the larger the range of initial distances between the star and the planet that leads to an engulfment.}.

The decrease of luminosity at the tip comes from the fact that the mass of the core at He-ignition in the 2 M$_{\odot}$ with $\Omega_{ini}/\Omega_{crit}=0.5$ is smaller than in the same model with $\Omega_{ini}/\Omega_{crit}=0.1$. At first sight, it might appear strange that the core mass is smaller in the faster rotating model. Indeed, rotation, in general, makes the masses of the cores larger. The point to keep in mind here is that we speak about the mass of the core required to reach a given temperature, which is the temperature for helium ignition. This mass depends
on the equation of state. In semi-degenerate conditions, the core mass needed to reach that temperature is larger than in non-degenerate ones  \citep[see e.g. Fig. 14 in][]{MM1989}. 
For a given initial mass, faster rotation, by increasing the core mass during the core H-burning,
makes the helium core less sensitive to degeneracy effects. 
In other words, rotation shifts to lower values the mass transition between He-ignition in semi-degenerate and non-degenerate regimes.

The evolution of the radius as a function of time along the RG branch in the case of the 2 M$_{\odot}$ models is shown in Fig. \ref{fig:struc} (left panel). 
The radius at the tip of the RG branch is lower by about 0.1 dex therefore by 25\% in the case of the faster rotating model. The small bump seen along the curves at time coordinates 0.06 and 0.075 is due to the fact that, when the star climbs the RG branch, the H-burning shell moves outwards in mass and at a given point encounters the chemical discontinuity left by the convective envelope that also slowly recedes outwards after the first dredge-up. This produces a rapid increase in the abundance of hydrogen in the H-burning shell, a variation in the
energy output of this shell and a change of the structure that produces those bumps.

The right panels of Fig. \ref{fig:struc} show how the masses and the radii at the base of the convective envelope vary as a function of time in both rotating models. In the faster rotating model, the convective envelope has a slightly smaller extension in mass and radius than in the slower rotating model. However these changes are minor.

From this brief discussion, we can conclude that rotation, by changing the transition mass between stars going through a He-flash and those avoiding it, may have a non negligible impact on the orbital evolution in this mass range. Outside this mass range, the impact of rotation will be modest.

\section{Planetary orbit evolution and stellar rotation}\label{sec:4}

\subsection{Planetary orbit evolution}

\begin{figure*}
\centering
\includegraphics[width=.49\textwidth, angle=0]{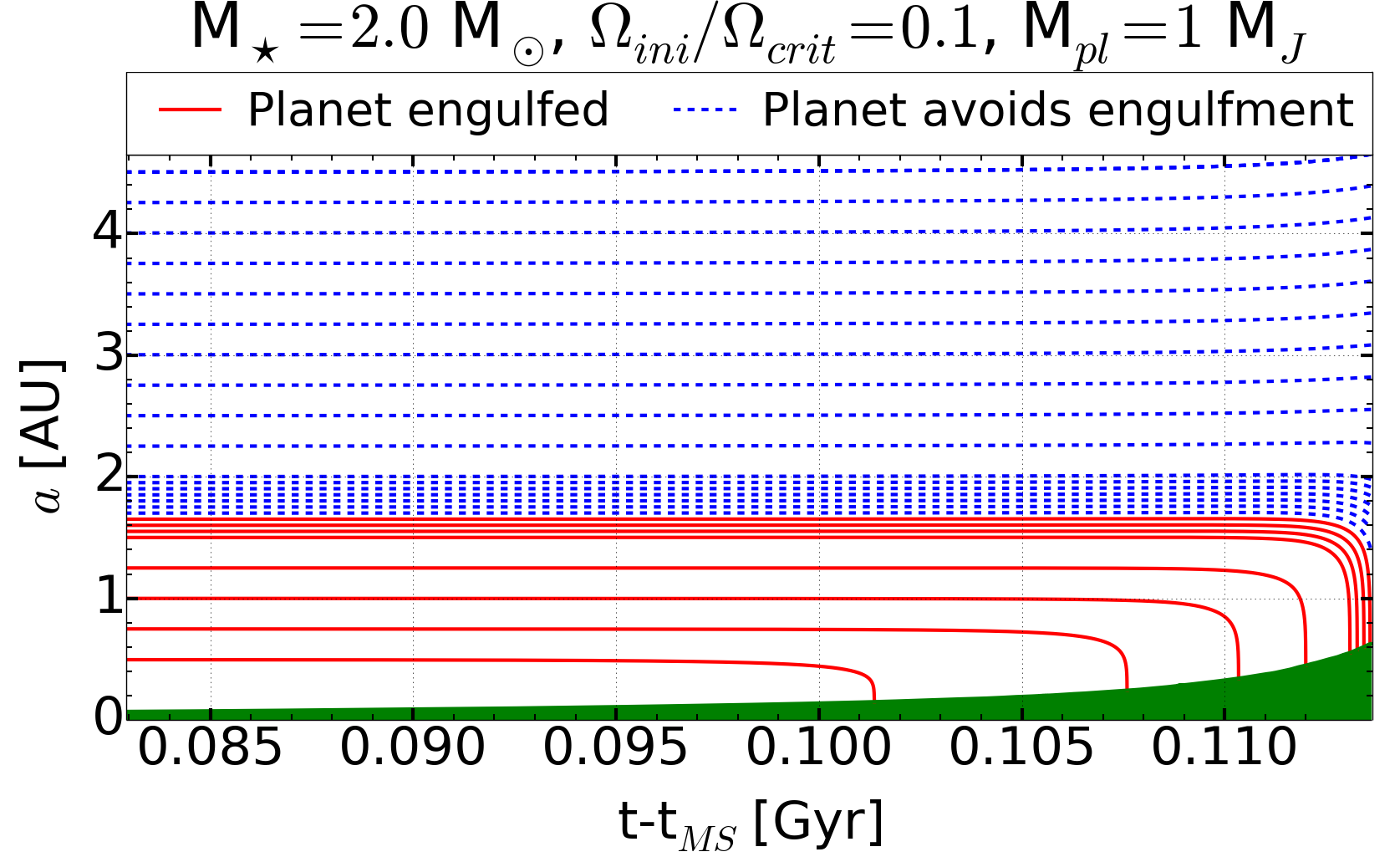}\includegraphics[width=.49\textwidth, angle=0]{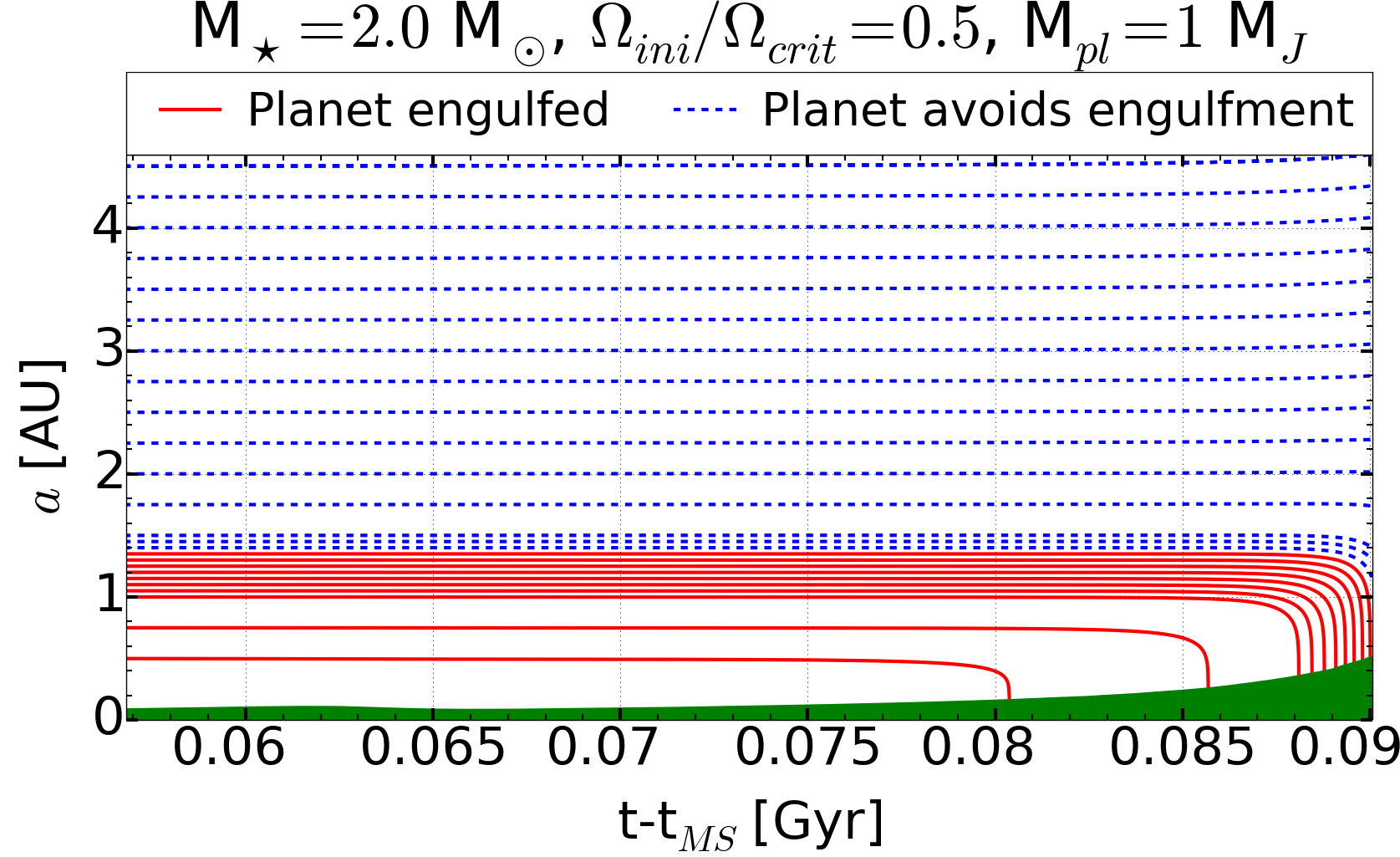}
\caption{
{\it Left panel:} Evolution of the semi-major axis of a 1 M$_{J}$ planet orbiting a 2 M$ _{\odot}$ star computed with an initial rotation on the ZAMS of $\Omega_{ini}/\Omega_{crit} = 0.1$. The different lines correspond to different initial semi-major axis values. Only the evolutions in the last 50-60 million years are shown. Before that time, the semi-major axis remains constant. The red solid lines represent the planets that will be engulfed before the star has reached the tip of the RG branch; the blue dashed lines represent the planet that will avoid the engulfment during the RG branch ascent. The upper envelope of the green area gives the value of the stellar radius.
{\it Right panel:} Same as the left panel, but for an initial rotation $\Omega_{ini}/\Omega_{crit} = 0.5$.}
\label{fig:orb15ms}
\end{figure*}

In Fig. \ref{fig:orb15ms}, we compare the evolution of the semi-major axis for planets of 1 M$_{J}$ orbiting around 2 M$_{\odot}$ stars with an initial slow and rapid rotation on the ZAMS. As previously obtained by many authors \citep[see {\it e.g.}][]{kunimoto11, villaver09, villaver14}, and as recalled in the previous section, the evolution of the orbit for planets that are engulfed is a kind of runaway process. Once the tidal forces begin to play an important role, a rapid decrease of the radius of the orbit is observed as a result of the very strong dependency of the tidal force on the ratio between the stellar radius that is increasing and the semi-major axis that is decreasing (term in $(R_*/a)^{-8}$ in Eq.~\ref{equa:tides}). Comparing the left and the right panel of Fig. \ref{fig:orb15ms}, we see that the interval of initial distances leading to an engulfment during the RG branch is a bit smaller for the faster rotating models (see the change of the limit between the continuous red and the dashed blue lines). As explained in Sect.~3 above, the larger the initial rotation rate, the lower the lower the luminosity at the RG tip. A lower luminosity leads to a lower value for the maximum stellar radius, and thus to more restricted conditions for having an engulfment.

\begin{figure*}
\includegraphics[width=.49\textwidth, angle=0]{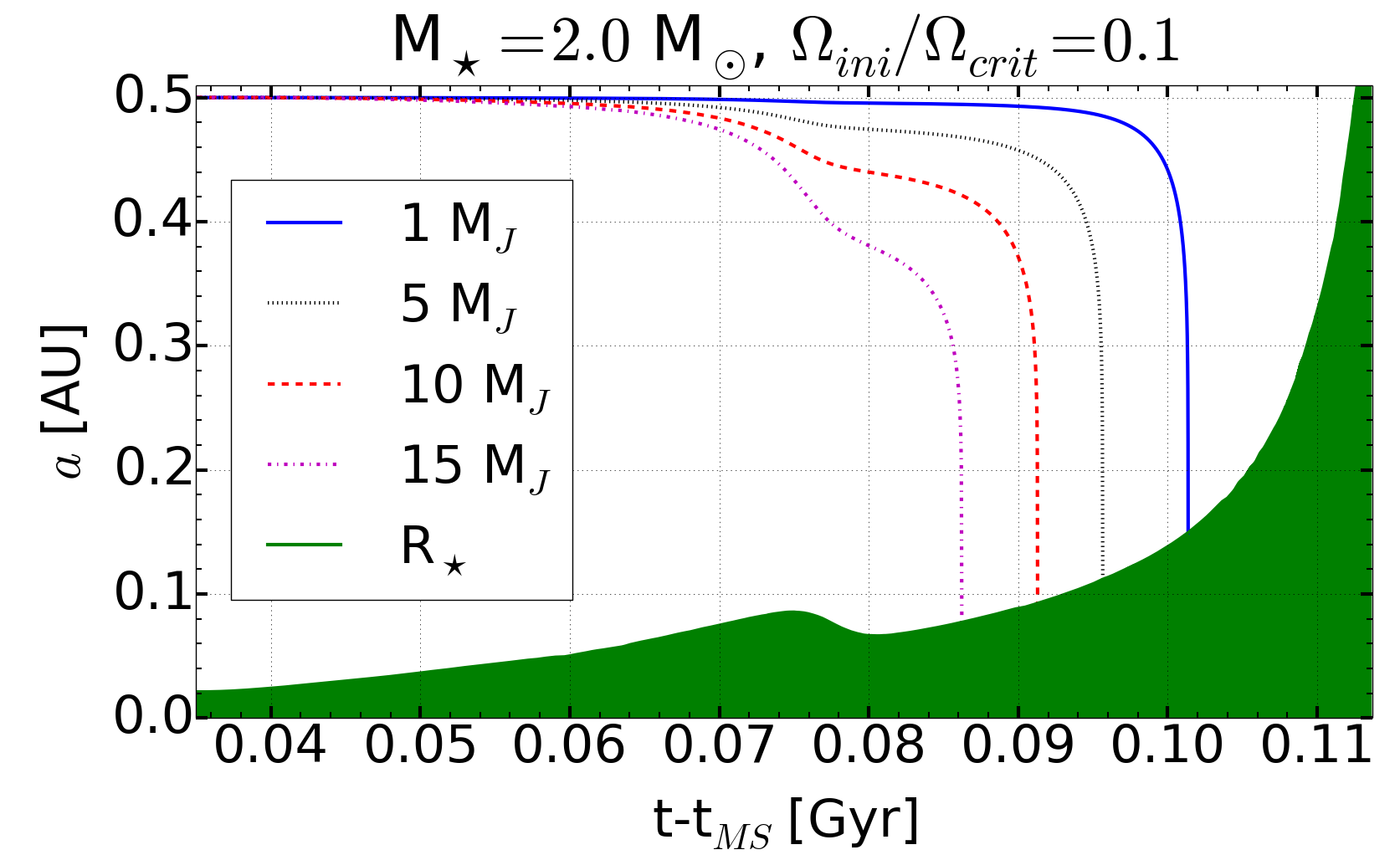}\includegraphics[width=.49\textwidth, angle=0]{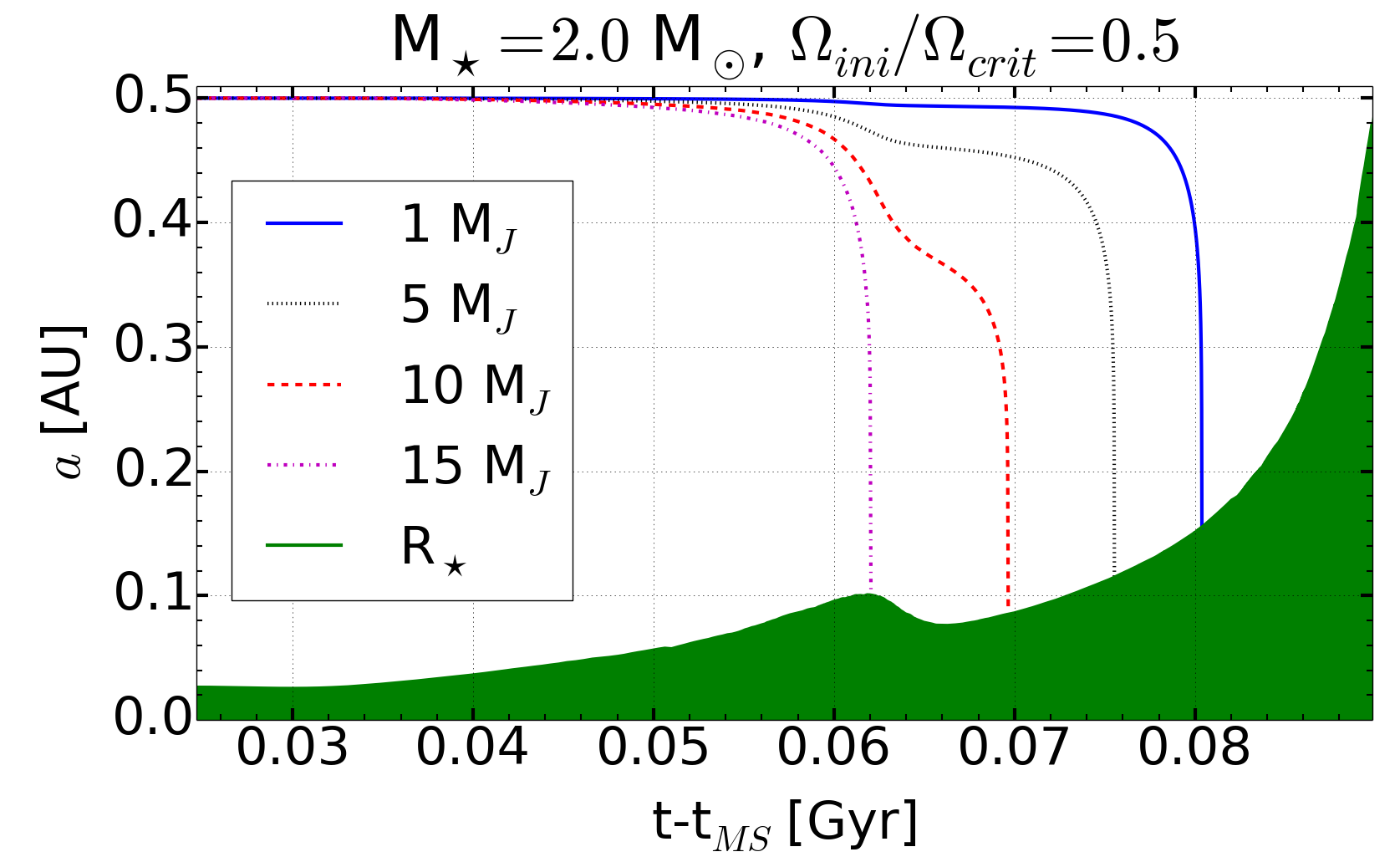}
\caption{{\it Left panel:} Evolution of the orbit for planets of masses equal to 1 M$_{J}$ (blue solid line), 5 M$_{J}$ (black dotted line), 10 M$_{J}$ (red dashed line) and 15 M$_{J}$ (magenta dashed-dotted line) around a 2 M$_{\odot}$ star with an initial rotation
equal to 10\% the critical angular velocity. The initial semi-major axis is equal to 0.5 au. {\it Right panel:} Same as the left panel, but for an initial rotation rate equal to 50\% the critical angular velocity.}
\label{migr}
\end{figure*}

Figure~\ref{migr} shows the impact of different initial rotation rates for a 2 M$_\odot$ on the orbital decay of planets of various masses beginning all their evolution at a distance equal to 0.5 au. The orbit for the 15 M$_J$ planet for instance
presents a different evolution around the slow and the fast rotator. 
The orbital decay occurring around the slow rotating star (magenta dashed-dotted line on the left panel)
occurs just during the bump. The decrease of the semi-major axis slows down when the stellar radius decreases
since the tidal force depends on $(R_*/a)^{-8}$.
Around the fast rotating model (see the magenta dashed-dotted line on the right panel), the orbit decay occurs just
at the beginning of the bump and is quite rapid.
As mentioned above, this illustrates that stellar rotation changes the structure of the star and modifies thereby the evolution of the orbit. In general, the engulfment occurs at an earlier evolutionary stage when the initial rotation rate increases.

\subsection{Impact of stellar rotation on the conditions leading to an engulfment}

\begin{figure}
\includegraphics[width=.49\textwidth, angle=0]{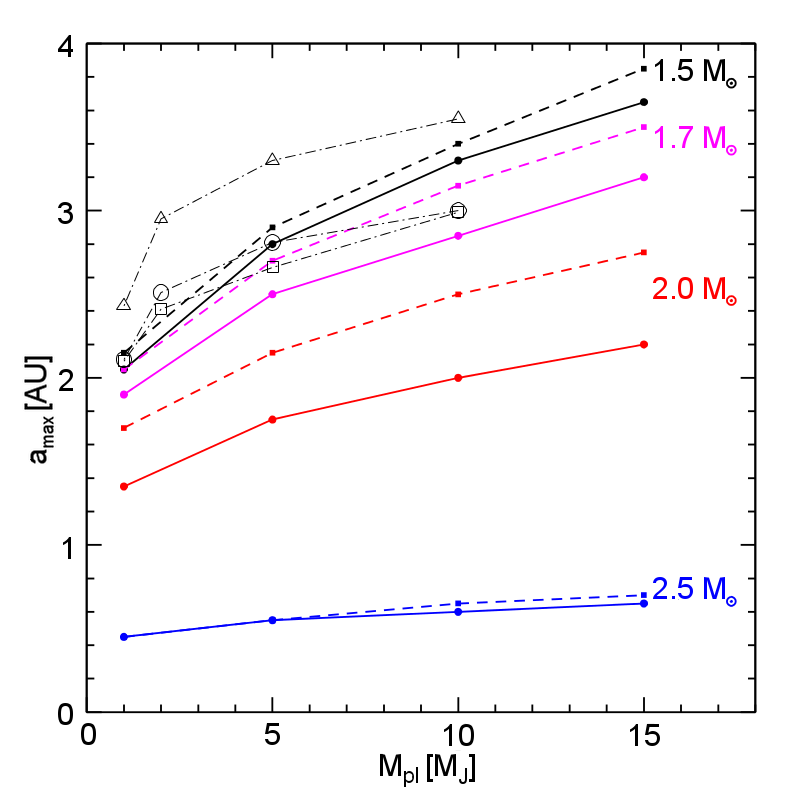}
\caption{Variation of the maximum semi-major axis below which engulfment occurs during the RG ascent as a function of the mass of the planet, of the mass of the star and its initial rotation rate.
The dashed and continuous lines are respectively for an initial stellar angular velocity equal to 10\% and 50\% the critical one. The empty symbols connected by light dashed-dotted lines are the results obtained by \protect\citet{villaver14} for planets
with masses of 1, 2, 5 and 10 M$_{J}$ orbiting a non-rotating 1.5 M$_{\odot}$ star. The triangles are for models using weak mass loss rates during the red giant phase ($\eta=0.2$ in Eq.~1), the circles are for models with normal red giant
mass losses  ($\eta=0.5$), and the squares are for models with $\eta=0.5$ and overshooting.}
\label{amax}
\end{figure}

As already discussed in the literature \citep[see {\it e.g.}][]{villaver09, villaver14}, we see that for a given mass of the planet and for given properties of the host star, there exists a maximum initial semi-major axis below which engulfment will occur during the RG phase. These maximal values are given in Table \ref{table:max_value_eng} and shown in Fig.~\ref{amax}. These results illustrate that the conditions for engulfment are more restricted around fast rotating stars, but the effect remains quite modest for the 1.5 and 2.5 M$_{\odot}$ models, while it is more important for the intermediate mass cases for the reasons already explained above. 

\begin{table}
\scriptsize{
\caption{Initial semi major axes below which the planet is engulfed during the RG phase.} 
\begin{tabular}{ccc}
\hline\hline 
\multicolumn{3}{c}{1.5 $M_{\odot}$, Z=0.02}\\
\hline 
$M_{pl}$ & $ a_{max}$ &    $ a_{max}$    \\
& ($\Omega_{ini}$/ $\Omega_{crit}$=0.1) &    ($\Omega_{ini}$/ $\Omega_{crit}$=0.5)    \\
\hline 
1    &    2.2    &    2.1    \\
5    &    2.9    &    2.8    \\
10    &    3.4    &    3.3    \\
15    &    3.85    &    3.65    \\
\hline 
\multicolumn{3}{c}{1.7 $M_{\odot}$, Z=0.02}\\
\hline 
$M_{pl}$ & $ a_{max}$ &    $ a_{max}$    \\
& ($\Omega_{ini}$/ $\Omega_{crit}$=0.1) &    ($\Omega_{ini}$/ $\Omega_{crit}$=0.5)    \\
\hline 
1    &    2.05    &    1.9    \\
5    &    2.7    &    2.45    \\
10    &    3.15    &    2.85    \\
15    &    3.5    &    3.15    \\
\hline 
\multicolumn{3}{c}{2.0 $M_{\odot}$, Z=0.02}\\
\hline 
$M_{pl}$ & $ a_{max}$ &    $ a_{max}$    \\
& ($\Omega_{ini}$/ $\Omega_{crit}$=0.1) &    ($\Omega_{ini}$/ $\Omega_{crit}$=0.5)    \\
\hline 
1    &    1.7    &    1.4    \\
5    &    2.15    &    1.75    \\
10    &    2.5    &    2    \\
15    &    2.75    &    2.2    \\
\hline 
\multicolumn{3}{c}{2.5 $M_{\odot}$, Z=0.02}\\
\hline 
$M_{pl}$ & $ a_{max}$ &    $ a_{max}$    \\
& ($\Omega_{ini}$/ $\Omega_{crit}$=0.1) &    ($\Omega_{ini}$/ $\Omega_{crit}$=0.5)    \\
\hline 
1    &    0.45    &    0.45    \\
5    &    0.55    &    0.55    \\
10    &    0.65    &    0.6    \\
15    &    0.7    &    0.65    \\
\hline 
\label{table:max_value_eng}
\end{tabular}
}
\end{table}

In Fig.~\ref{amax}, we note that the effects of rotation do not produce overlaps between the curves, indicating that the change due to the initial mass dominates over the change due to rotation (at least for the range of rotation rates considered here). The present results for the evolution of planetary orbits are well in line with previous works \citep{kunimoto11, villaver09, villaver14}. More specifically, we also obtain that the conditions for engulfment are more favorable for more massive planets and less massive stars (note that this is because the less massive stars reach larger luminosities at the tip of the RGB). Moreover, \citet{kunimoto11} found that the orbital radius above which planet engulfment is avoided is quite sensitive to the stellar mass at the transition between those going through a helium flash and those avoiding the helium flash. Qualitatively, this is exactly what we find here (see Fig. \ref{amax}). 

\begin{figure}
\includegraphics[width=.49\textwidth, angle=0]{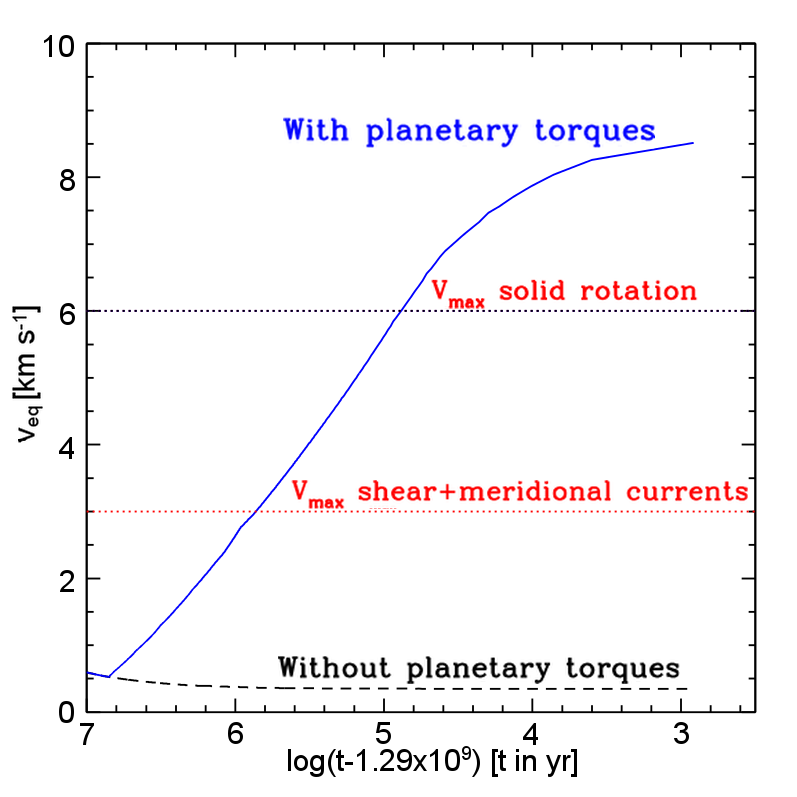}
\caption{Evolution of the equatorial velocity at the surface of a 2 M$_{\odot}$. The dashed line corresponds to the evolution obtained from a model with an initial surface angular velocity equal to 10\% the critical
angular velocity without tidal interaction. The continuous line corresponds to the same model but accounting for the transfer of angular momentum from the planetary orbit to the star (15 M$_{J}$ planet beginning to orbit at 1 au on the ZAMS).  
The line stops at engulfment. The two horizontal dotted lines 
indicate the maximum velocities that can be produced in some extreme situations from single stars at the position in the HR diagram where engulfment occurs. The lowest limit corresponds to models computed with the same physics of rotation as here but starting from a much higher initial rotation
equal to about 95\% the critical rotation on the ZAMS. The upper limit would correspond to a model starting with the same very high initial velocity assuming solid-body rotation during its whole evolution.}
\label{vitstar1}
\end{figure}

\begin{figure}
\includegraphics[width=.49\textwidth, angle=0]{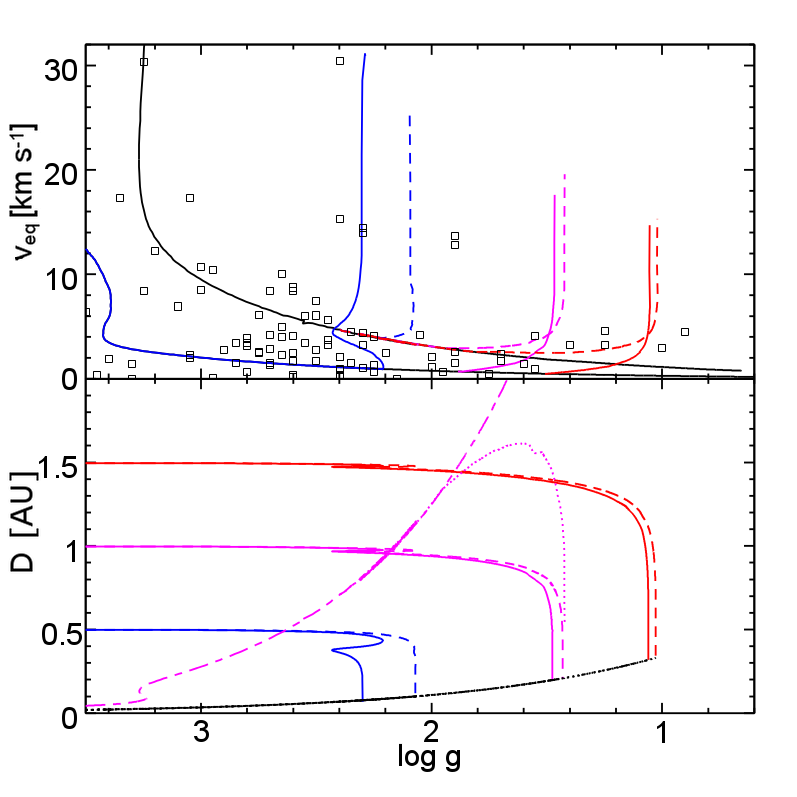}
\caption{
The upper panel shows the evolution of the surface equatorial velocity for various 2 M$_\odot$ models with and without planets. The continuous black curves are for models without planets. The lower curve is for the models with
$\Omega_{\rm ini}/\Omega_{\rm crit}=0.1$, the upper curve for $\Omega_{\rm ini}/\Omega_{\rm crit}=0.5$. The colored curves are for models with planets. The evolution is shown only until the engulfment. The continuous (dashed) curves
correspond to $\Omega_{\rm ini}/\Omega_{\rm crit}=0.1$ ($\Omega_{\rm ini}/\Omega_{\rm crit}=0.5$) models. The planets have a mass equal to 15 M$_J$. From left to right, are shown the case for
an initial distance between the star and the planet equal to 0.5, 1.0 and 1.5 au. The lower panel shows the evolution of the semi-major axis of the planetary orbit (colored continuous and dashed curves) and the evolution of the stellar radii (black lower curves).
The long-short dashed (magenta) curve shows the evolution of the corotation radius for the 2 M$_\odot$ with $\Omega_{\rm ini}/\Omega_{\rm crit}=0.5$ without planet, the dotted line for the same model with a tidal interaction with a 15 M$_j$ planet
beginning its orbital evolution at a distance of 1 au from the star. The curve stops at engulfment.
}
\label{vitstar2}
\end{figure}

\begin{figure}
\includegraphics[width=.49\textwidth, angle=0]{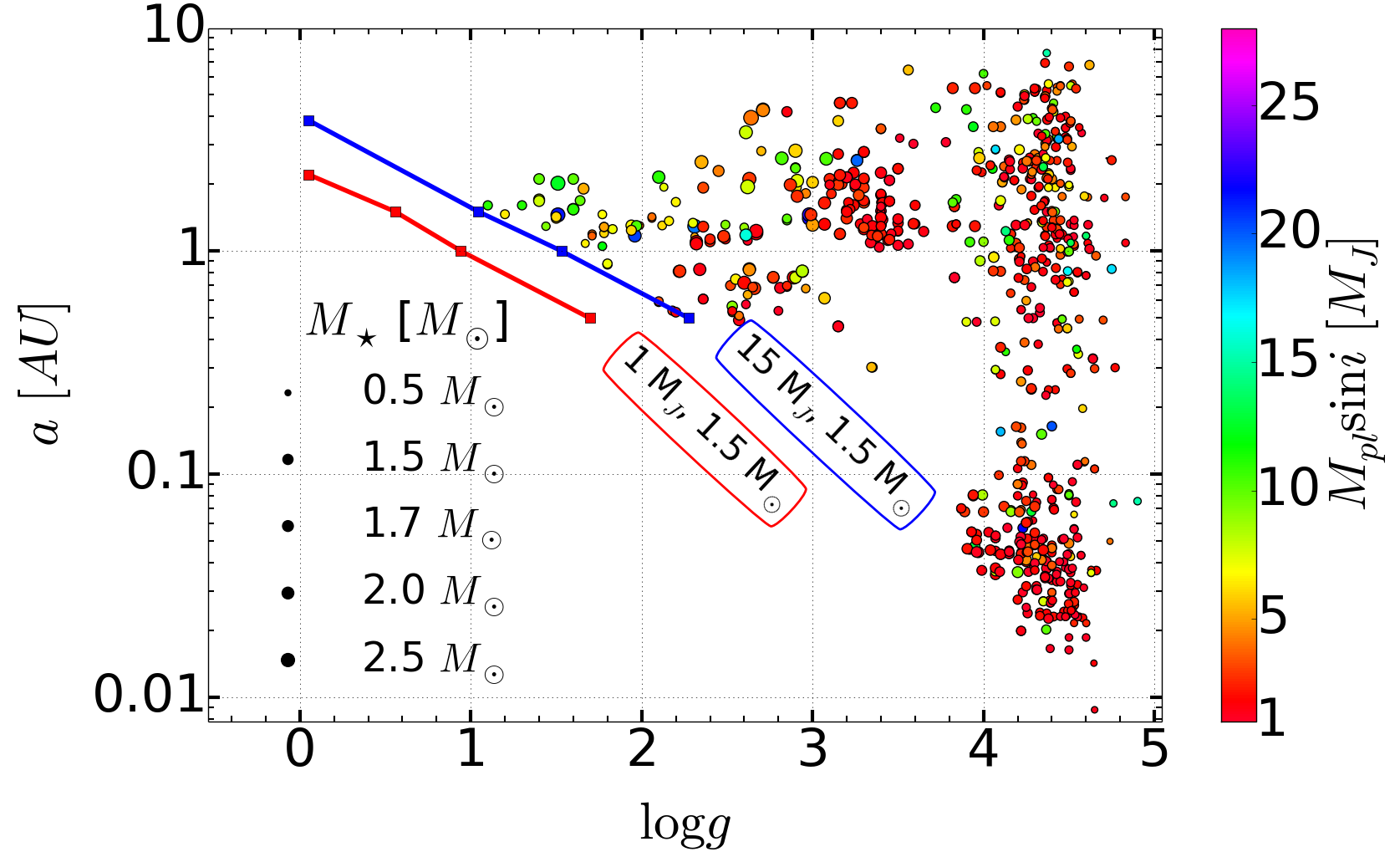}
\caption{Semi-major axis of planetary orbits versus the surface gravity (log $g$) of the host stars using the database of the exoplanets.org \protect\citep{Han2014} and of NASA Exoplanet Archive \protect\citep{akeson13}. The colors allow to have an indication on the mass of the planet, while the size of the symbols are related to the mass of the star. The thick lines in the upper left corner show the minimum semi-major axis for survival of the planet as computed in the present work for stars with $\Omega_{\rm ini}/\Omega_{\rm crit}=0.10$. The case for $\Omega_{\rm ini}/\Omega_{\rm crit}=0.50$ would be hardly distinguishable.
These lines are labeled by the masses of the planet and of the star considered.}
\label{amaxo}
\end{figure}

In Fig.~\ref{amax}, we have also plotted the semi-major axis above which no engulfment occurs (survival limit) predicted by \citet{villaver14} for planets with masses between 1, 2, 5 and 10 M$_{J}$ around a non-rotating 1.5 M$_{\odot}$ computed with different physical ingredients (see caption). The upper curve from \citet{villaver14} was obtained with a smaller mass loss rate during the ascent of the RG branch. Lowering the mass loss rate leads to larger radii at the tip of the RG branch and thus shifts the survival limit to larger values. The two lower curves by \citet{villaver14} use the same mass loss rates as in the present work. The two curves differ by the inclusion of overshooting, the lower curve (empty squares) being the one computed with overshooting.  This last model would likely be the one that is the most comparable with the present models although there are differences (for instance the chemical composition is different). Anyway, the differences are small and we note that increasing the core size shifts a bit the curve downward as in the present models. Interestingly, we note that mass losses along the RG branch has likely a larger effect than rotation for this specific initial mass.

\subsection{Impact of the orbital decay on the stellar rotation}

When the semi-major axis of the planetary orbit is inferior to the corotation radius, 
the tidal forces can transfer angular momentum from the orbit of the planet to the star, accelerating its rotation.
The question we want to address here is whether this would produce some observational effects before the engulfment\footnote{The case of the engulfment is treated in the second paper of the series.}. We can wonder whether such a transfer can lead to high surface velocities at the stellar surface that would be difficult to explain by any other mechanism. If yes, we shall then have to investigate whether these velocities will be reached for a sufficiently long period for being observable.
Let us begin by a few orders of magnitude considerations. From Tables \hyperref[table:ap1]{A.1} and \hyperref[table:ap2]{A.2}, we see that the angular momentum contained in the orbit ranges between 1 and 46 times the angular momentum contained in the star (see third column). Thus we can expect that transfer of even half of this angular momentum will have a significant impact on the angular momentum of the star. Due to the orbital decay, the planet will lose a significant part of its angular momentum (see the fraction of the initial orbital angular momentum lost by the orbital decay in column 9). At the moment of engulfment, the planet will have therefore an angular momentum which is only a fraction of its initial value (the angular momentum at time of engulfment is given in column 8). Of course the sum of the fraction given in columns 8 and 9 gives 1. 

Only part of the orbital angular momentum lost by the planet is transferred to the star, the part due to the tidal forces. The part of the orbital decay due to the other processes (stellar winds, changes of the planet mass, friction and gravitational drag forces) do not transfer any angular momentum from or to the star (the fraction of the angular momentum lost by these processes is indicated in column 10 as a fraction of column 9). We then obtain the fraction (in percents) of the initial orbital angular momentum transferred to the star by
 $\mathcal{L}_{migr}(1-f_{notides}/100)  \sim \mathcal{L}_{migr}$. This fraction is in general more than half of the initial orbital angular momentum in agreement with previous estimates  \citep{carlberg09}.

In the case of the 2 M$_{\odot}$ star with $\Omega/\Omega_{crit}=0.1$ and a 15 M$_{J}$ planet beginning to orbit at 1 au on the ZAMS,  the angular momentum transferred from the orbit to the star is equivalent to more than 22 times the angular momentum contained in the convective outer envelope. This implies that
the angular velocity of the star will be enhanced by a factor of about 23 as well as the surface velocity, that will increase from 0.47 km s$^{-1}$ to 10.8 km s$^{-1}$.

Actually, this process does not occur instantaneously and during the transfer the structure of the star changes and thus the real increase will be less than the one obtained from this simple estimate.

In Fig.~\ref{vitstar1}, we show how the surface velocity of the star increases due to the orbital decay in the case of the 2 M$_\odot$ with $\Omega/\Omega_{crit}=0.1$ and a 15 M$_{J}$ planet beginning to orbit at 1 au on the ZAMS. We see that the whole acceleration process takes place in less than 10 Myr. This duration corresponds to about 17\% of the RG branch ascent phase, which is not negligible, and indicates that it is not unreasonable to observe systems which would be in such a phase. Before the engulfment, the period during which the surface velocity would be larger than 6 and 8 km s$^{-1}$ would be respectively two and three orders of magnitude shorter ({\it i.e.} of the order of 100 000 yr and 10 000 yr, respectively). The duration of these phases of high surface rotation rates {\it before engulfment}, although not negligible, are respectively only 2 thousandths or 2 ten thousandths of the RG branch duration, thus are much more difficult to observe. As we shall see in the second paper of this series, these high velocities will be maintained for long periods {\it after engulfment}, indicating that high velocity red giants will be more easily observed after engulfment than just before.

In Fig.~\ref{vitstar1}, we have also indicated the values of the surface velocity reached by a 2 M$_{\odot}$ star at the same position in the HR diagram as where the engulfment occurs, but assuming that the star started with a very high initial rotation rate on the ZAMS equal to 95\% the critical velocity. The star is assumed to evolve as a single star without planet engulfment or any other interaction. We see that such an evolution cannot predict surface rotation rates beyond a velocity of about 3 km s$^{-1}$. 
Of course the surface velocity depends on the way angular momentum is transported inside the star. The values shown in Fig.~\ref{vitstar1} have been obtained with the physics presented in Sect. \ref{sec:2}. Would we have assume that stars rotate as solid bodies, then one would obtain the limit shown by the upper dotted line. Even in that case, the velocities remain below 6 km s$^{-1}$. This indicates that surface velocities for red giants above a value of 6 km s$^{-1}$ and for a surface gravity log $g\sim$ 1.5 cannot be obtained for single stars. Such a high surface velocity for a red giant with a low surface gravity is the signature of an interaction. 

We shall end this section by comparing the kinetic energy of the planet just before engulfment with the binding energy of the convective envelope (see columns 11 and 13 in Tables \hyperref[table:ap1]{A.1} and \hyperref[table:ap2]{A.2}). We see that the kinetic energy is two to three orders of magnitude smaller than the binding energy, thus it is not expected that any striping off the envelope will occur as a result of injection into the convective envelope of the kinetic energy of the planetary orbit. Let us note also that the enhancement of the rotation rate at the surface will not be able to make the star to reach the critical velocity. Indeed, we just saw above that the velocity will reach values of the order of 10 km s$^{-1}$, while the critical velocity along the RG branch for a 2 M$_{\odot}$ star is of the order of 100 km s$^{-1}$.

\subsection{Comparisons with observed systems}

In the upper panel of Fig.~\ref{vitstar2}, we compare the surface velocities obtained for our 2 M$_\odot$ models with and without tidal interactions with the observed sample of red giant stars observed by \citet{carlberg12}. 
The results of the present work show that the most clear candidates for a tidal interaction (see the points above the track for the fast rotating single star model) are those red giants located near the end of the RG branch, {\it i.e.} for log $g < 2.5$. The velocities obtained by models with tidal interactions can reach values equal or even above those that would be needed to reproduce these fast rotating red giants. While some
of these red giants might still be in a stage before an engulfment (for instance those with $\upsilon\sin i = 3-5$ km s$^{-1}$ at log $g$ below 1.4), others (those with $\upsilon\sin i$ around 14 km s$^{-1}$ with log $g$ between 1.9 and 2.4) are more likely red giants having evolved after a tidal interaction/engulfment. Indeed, the duration of the rapid increase of the surface velocity before the engulfment is so short that there is little chance to observe stars at that stage; there is however more chance to observe stars after the engulfment. 

In the lower panel of Fig.~\ref{vitstar2}, the evolution of the corresponding orbits of 15 M$_{J}$ planets are shown. The evolution of the corotation distance is also shown 
for a 2 M$_\odot$ model with (magenta dotted line) and without a planet (magenta short-long dashed curve). The corotation distance decreases when the star accelerates as a result of the tidal acceleration (see the dotted magenta line in Fig.~\ref{vitstar2}), however, the orbital radius always remains interior to the corotation radius and thus an engulfment occurs.

Another interesting plot to check whether the theoretical predictions are compatible with current observations is the one shown in Fig. \ref{amaxo}, where the semi-major axis of planets of various masses are plotted as a function of the gravity of the star.
The most striking fact is that one can observe planets with much smaller semi-major axis around main-sequence stars, {\it i.e.} with log $g$ larger than 4, than around evolved stars, {\it i.e.} with log $g$ smaller than about 3.0 \citep[see {\it e.g.}][]{kunimoto11}. Typically, for stars with log $g$ smaller than about 3.0, no planets with a semi-major axis inferior to about 0.5 au are observed. This might reflect the fact that planets with smaller semi-major axis are engulfed by the star. 

From the present computations, we can plot on that diagram the gravity at which engulfment occurs on the RG branch for different initial semi-major axis. In the case of 1 M$_{J}$ planets around 1.5 M$_{\odot}$ stars, we obtain the thick red lines. For 15 M$_{J}$ planets, we obtain the thick blue line. The results are obtained for the initially slow rotating stellar models. The case for the fast rotating model would be hardly distinguishable in Fig. \ref{amaxo}. According to present computations, one should hardly observe any systems with smaller semi-major axis than those given by these lines at the considered gravities. We see that indeed, for the systems considered in this work, all observed systems have semi-major axis which are larger than the predicted limits. 

Some empty gap however seems to remain between the theoretical limits obtained here and the observations. At the moment, it is difficult to know whether such a gap really exists or not, since a careful study of the observational biases would be needed.
Such a gap, if real,  might indicate that the true survival limit would still be larger than the one predicted here.

We end this section with a few comments about the frequency of planet engulfments. Let us first consider the case of 1 M$_{J}$ planets around 2 M$_{\odot}$ stars. If we assume that the survival limit is 1.5 au (an intermediate value between 1.4 and 1.7 au indicated in Table \ref{table:max_value_eng}), that there are Jupiter like planets around a fraction $f$ of stars, and that the probability for these planets to be at distance $a$ and $a+{\rm d}a$ from its parent star is given by $2\pi a {\rm d}a/\pi a^2_{max}$ with $a_{max}$ the maximum initial radius of the orbits of such planets, one obtains that the probability of planet engulfment would be $f a_{survival}^2/a_{max}^2$. If we set $f\sim 1$, $a_{survival}=1.5$ and $a_{max}=10$, one obtains a probability equal to 2.25\%. The hypotheses done are very schematic and are likely not very realistic. It is even difficult to say whether the result obtained is an upper or a lower limit. Nevertheless, it gives already an interesting information, namely that if the distances at birth are uniformly distributed, then one expects that about one percent of 2 M$_{\odot}$ red giants would engulf a planet. This number will be slightly larger for lower-mass stars and slightly smaller for more massive stars.

\section{Conclusions}\label{sect5}

In the present paper, we studied the evolution of the orbits of planets with masses between 1 and 15 M$_{J}$ around stars with masses
between 1.5 and 2.5 M$_{\odot}$. 

The originality of the present work lies in the fact that we used rotating stellar models allowing us to study for the first time the impact of the stellar rotation on the evolution of the planetary orbit, as well as the feedback of the evolution of the planetary orbit onto the rotation of the star itself.
The main results are the following:
\begin{itemize}
\item The present rotating stellar models allow to study the range of surface rotations expected along the RG branch
for stars evolving without tidal interactions. For stars with initial masses below 2.5 M$_\odot$, whatever the initial rotation, the surface velocities
are smaller than 5 km s$^{-1}$ for surface gravities in log $g$ below 2. Note that the first dredge-up does not produce any significant surface acceleration.
\item The surface velocity limit separating the normal rotating red giants (no tidal interaction) from red giants whose surface velocities, for being explained, need some acceleration mechanism depends on the mass of the star and on its surface gravity. For instance, for a 2.5 M$_\odot$ star, this limit is around 15 km s$^{-1}$
for a log $g$ equal to 2.6 and equal to 2 km s$^{-1}$ for a log $g$ equal to 1.
\item The best candidates of red giants having been accelerated by tidal interactions with a companion (and possibly also by an engulfment of a companion) have to be searched
in the upper part of the RG branch, where stellar models predict surface velocities below 5 km s$^{-1}$ even starting from very high initial rotation rates.
\item The orbital decay occurs at earlier evolutionary stages in faster rotating models. This is a consequence of the changes of the structure of the star due to rotation, in particular the fact that rotating stars are slightly overluminous and thus have a larger radius at a given effective temperature.
\item The survival domain of planets around stars initially rotating fast are more restricted than the survival domain of planets around stars initially rotating slowly.
This survival limit is the most sensitive to the initial rotation in the mass range around the mass for He-flash
(this transition is between 2 and 2.5 M$_\odot$ according to the present models). In this domain, passing from a low initial angular velocity equal to 10\% the critical value to 50\% decreases the minimum semi-major axis for survival by about 20\%. Outside this mass range, the impact of rotation is more modest. 
Note that in the helium flash transition domain, the survival limit depends also sensitively on other parameters like {\it e.g} the core overshooting or the metallicity. 
\item The surface rotation of the star begins to increase before the engulfment (typically a few 10$^5$ years before in the case of our 2 M$_\odot$  model with $\Omega_{\rm ini}/\Omega_{\rm crit}=0.1$ and a 15 M$_{J}$ planet beginning to orbit at 1 au on the ZAMS) and can 
be enhanced by factors between 10 and 20. These velocities are still well below the critical velocity and well below the value that would be needed for the corotation radius to be smaller than the actual radius of the orbit, and thus for making the tidal forces to reverse their direction and the planet to follow an outward migration. High surface velocities (typically higher than 10 km s$^{-1}$) will be reached only during very short periods before engulfment. 
As we shall see in the second paper of this series, pursuing the evolution of the star beyond engulfment shows that
the high velocities reached by tidal interactions and by the engulfment itself will not disappear in short timescales and can produce fast rotating red giants.

\end{itemize}

\begin{acknowledgements} 
This research has made use of the Exoplanet Orbit Database
and the Exoplanet Data Explorer at exoplanets.org. The project has been supported by Swiss National Science Foundation grants 200021-138016 and  200020-15710.
AAV acknowledges support from an Ambizione Fellowship of the Swiss National Science Foundation.
\end{acknowledgements}

\begin{appendix}

\section{Some properties of the planetary orbits}\label{sec:appe}

In Tables \hyperref[table:ap1]{A.1} and \hyperref[table:ap2]{A.2}, the following properties of the orbits are indicated: Columns 1 to 3 indicate respectively the mass of the planet, the initial angular momentum in the planetary orbit and the ratio of the initial orbital angular momentum and the axial angular momentum of the star. Columns 4 to 7 show the duration of the planet migration, the age, the luminosity and gravity of the star when the engulfment occurs. The duration of the migration is the time between the stage when 10\% of the initial angular momentum of the planet has been lost and the time of engulfment. The quantity of angular momentum that will be given to the star at the moment of engulfment is indicated in column 8, the angular momentum lost by the planet during the orbital decay is given in column 9. The fraction of this orbital angular momentum lost during the orbital decay that is due to the forces other than tides is indicated in column 10. Note that the angular momentum transferred from the planetary orbit to the star is given by column 9 multiplied by [1-(column 10)/100]. Indeed, the forces other than the tides transfer angular momentum to the circumplanetary material not to the star. Finally columns 11 to 13 show the kinetic energy of the planet at the engulfment, the angular momentum in the external convective envelope of the star at the engulfment and the potential energy of the stellar convective envelope at engulfment.

\begin{table*}
\scriptsize{
\caption{Various properties (see Appendix \hyperref[sec:appe]{A}) of the planetary orbits with planet host stars having an initial angular surface velocity equal to 10\% the critical angular velocity.} 
\resizebox{18cm}{!} {
\begin{tabular}{ccccccccccccc}
\hline\hline 
$M_{pl}$ & $\mathcal{L}_{0,pl}$  & $\mathcal{L}_{0,pl}/\mathcal{L}_{0,\star}$ &$t_{migr}$ & Age$_{eng}$& $\log L_{eng}/L_{\odot}$& $\log g_{eng} $ & $\mathcal{L}_{eng}$ &$\mathcal{L}_{migr}$   & $f_{notides}$ & $E_{kin}(Age_{eng})$ &  $\mathcal{L}_{env}(Age_{eng})$ & $E_{pot}(Age_{eng})$  \\ 
 $[M_{J}]$ & [$10^{50}$ g cm$^2$ s$^{-1}$] & &[Myr] & [Gyr]& & &[$\% \ \mathcal{L}_{0,pl}$]& [$\% \ \mathcal{L}_{0,pl}$]& [$\% \ \mathcal{L}_{migr}$]    & [$10^{44}$ erg] & [$\% \ \mathcal{L}_{0,\star}$]& [$10^{47}$ erg] \\
 (1)&(2)&(3)&(4)&(5)&(6)&(7)&(8)&(9)&(10)&(11)&(12)&(13) \\
 \hline 
 \multicolumn{13}{c}{ }\\
 \multicolumn{13}{c}{1.5 $M_{\odot}$, $\Omega_{ini}$/ $\Omega_{crit}$=0.1, $\mathcal{L}_{0,\star} = 0.41$ [$10^{50}$ g cm$^2$ s$^{-1}$], Z=0.02}\\
 \multicolumn{13}{c}{ }\\
\hline 
\multicolumn{13}{c}{a$_{ini}=0.5$ [au], $P\simeq0.29$ [yr]}\\
\hline 
1    &    0.73    &    1.80    &    0.96    &    3.445    &    2.289    & 1.696 &    40.46    &    59.54    &    0.97      &    0.25      &    75.38    &    -3.31    \\
5    &    3.67    &    8.99    &    2.98    &    3.436    &    2.059    & 1.971 &    29.79    &    70.21    &    3.08    &    1.26      &    77.61    &    -4.50    \\
10    &    7.35    &    18.01    &    7.88    &    3.428    &    1.922    & 2.141 &    27.23    &    72.77    &    5.05    &    2.51    &    78.48    &    -5.38    \\
15    &    11.04    &    27.06    &    18.12    &    3.420    &    1.826    & 2.275 &    22.37    &    77.63    &    6.66    &    3.75    &    78.95    &    -6.08    \\
\hline 
\multicolumn{13}{c}{a$_{ini}=1$ [au], $P\simeq0.82$ [yr]}\\
\hline 
1    &    1.04    &    2.54    &    0.32    &    3.457    &    2.843    & 0.942 &    52.88    &    47.12    &    0.91    &    0.13    &    60.64    &    -1.49    \\
5    &    5.19    &    12.72    &    0.91    &    3.454    &    2.637    & 1.210 &    42.22    &    57.78    &    3.01    &    0.63    &    68.63    &    -2.02    \\
10    &    10.39    &    25.47    &    2.30    &    3.452    &    2.512    & 1.360 &    30.51    &    69.49    &    4.93    &    1.26    &    71.76    &    -2.46    \\
15    &    15.62    &    38.27    &    4.73    &    3.449    &    2.422    & 1.530 &    27.90    &    72.10    &    6.48    &    1.88    &    73.45    &    -2.78    \\
\hline 
\multicolumn{13}{c}{a$_{ini}=1.5$ [au], $P\simeq1.5$ [yr]}\\
\hline 
1    &    1.27    &    3.11    &    0.22    &    3.460    &    3.151    & 0.557 &    57.77    &    42.23    &    0.96    &    0.08    &    37.41    &    -0.95    \\
5    &    6.35    &    15.58    &    0.55    &    3.458    &    2.940    & 0.827 &    39.79    &    60.21    &    3.03    &    0.42    &    55.01    &    -1.30    \\
10    &    12.73    &    31.20    &    1.26    &    3.457    &    2.827    & 0.942 &    35.17    &    64.83    &    4.93    &    0.84    &    61.47    &    -1.54    \\
15    &    19.12    &    46.87    &    2.36    &    3.456    &    2.745    & 1.044 &    31.09    &    68.91    &    6.42    &    1.25    &    65.06    &    -1.74    \\
\hline 
\multicolumn{13}{c}{ }\\
\multicolumn{13}{c}{1.7 $M_{\odot}$, $\Omega_{ini}$/ $\Omega_{crit}$=0.1, $\mathcal{L}_{0,\star} = 0.46$  [$10^{50}$ g cm$^2$ s$^{-1}$], Z=0.02}\\
\multicolumn{13}{c}{ }\\
\hline 
\multicolumn{13}{c}{a$_{ini}=0.5$ [au], $P\simeq0.27$ [yr]}\\
\hline 
1    &    0.78    &    1.70    &    0.80    &    2.206    &    2.334    &    1.705 &    42.28    &    57.72    &    0.76    &    0.29    &    81.06    &    -4.02    \\
5    &    3.91    &    8.49    &    1.95    &    2.198    &    2.120    &    1.976 &    30.71    &    69.29    &    2.38    &    1.43    &    82.83    &    -5.31    \\
10    &    7.82    &    17.00    &    4.15    &    2.192    &    2.003    &    2.133 &    26.48    &    73.52    &    3.89    &    2.85    &    83.51    &    -6.28    \\
15    &    11.75    &    25.54    &    7.63    &    2.187    &    1.921    &    2.233 &    24.42    &    75.58    &    5.11    &    4.26    &    83.86    &    -6.95    \\
\hline 
\multicolumn{13}{c}{a$_{ini}=1.0$ [au], $P\simeq0.77$ [yr]}\\
\hline 
1    &    1.10    &    2.40    &    0.31    &    2.217    &    2.884    &    0.931 &    50.88    &    49.12    &    0.77    &    0.14    &    68.49    &    -1.85    \\
5    &    5.52    &    12.01    &    0.71    &    2.214    &    2.690    &    1.231 &    39.70    &    60.30    &    2.51    &    0.71    &    75.10    &    -2.45    \\
10    &    11.06    &    24.05    &    1.48    &    2.212    &    2.576    &    1.383 &    32.20    &    67.80    &    4.11    &    1.42    &    77.63    &    -2.90    \\
15    &    16.61    &    36.12    &    2.67    &    2.210    &    2.495    &    1.512 &    28.91    &    71.09    &    5.41    &    2.13    &    79.00    &    -3.23    \\
\hline 
\multicolumn{13}{c}{a$_{ini}=1.5$ [au], $P\simeq1.41$ [yr]}\\
\hline 
1    &    1.35    &    2.94    &    0.20    &    2.220    &    3.190    &    0.489 &    54.34    &    45.66    &    0.81    &    0.10    &    48.12    &    -1.18    \\
5    &    6.76    &    14.70    &    0.44    &    2.218    &    2.989    &    0.804 &    44.25    &    55.75    &    2.60    &    0.48    &    63.13    &    -1.57    \\
10    &    13.55    &    29.45    &    0.89    &    2.217    &    2.883    &    0.931 &    35.92    &    64.08    &    4.22    &    0.95    &    68.53    &    -1.85    \\
15    &    20.35    &    44.24    &    1.50    &    2.216    &    2.808    &    1.040 &    33.03    &    66.97    &    5.52    &    1.42    &    71.50    &    -2.06    \\
\hline 
\multicolumn{13}{c}{ }\\
\multicolumn{13}{c}{2.0 $M_{\odot}$, $\Omega_{ini}$/ $\Omega_{crit}$=0.1, $\mathcal{L}_{0,\star} = 0.61$ [$10^{50}$ g cm$^2$ s$^{-1}$], Z=0.02}\\
\multicolumn{13}{c}{ }\\
\hline 
\multicolumn{13}{c}{a$_{ini}=0.5$ [au], $P\simeq0.25$ [yr]}\\
\hline 
1    &    0.85    &    1.40    &    0.60    &    1.286    &    2.402    &    1.733 &    44.66    &    55.34    &    0.52    &    0.34    &    86.63    &    -5.13    \\
5    &    4.23    &    7.00    &    1.24    &    1.280    &    2.202    &    1.997 &    39.21    &    60.79    &    1.59    &    1.68    &    87.80    &    -6.65    \\
10    &    8.48    &    14.02    &    3.10    &    1.276    &    2.081    &    2.145 &    28.38    &    71.62    &    2.55    &    3.35    &    88.27    &    -7.94    \\
15    &    12.73    &    21.05    &    9.24    &    1.271    &    1.954    &    2.301 &    24.19    &    75.81    &    3.15    &    5.02    &    88.58    &    -9.37    \\
\hline 
\multicolumn{13}{c}{a$_{ini}=1$ [au], $P\simeq0.71$ [yr]}\\
\hline 
1    &    1.20    &    1.98    &    0.25    &    1.295    &    2.939    &    1.014 &    54.44    &    45.56    &    0.58    &    0.17    &    77.28    &    -2.39    \\
5    &    5.99    &    9.90    &    0.49    &    1.293    &    2.763    &    1.240 &    40.73    &    59.27    &    1.89    &    0.84    &    82.04    &    -3.14    \\
10    &    11.99    &    19.82    &    0.81    &    1.291    &    2.663    &    1.411 &    34.44    &    65.56    &    3.08    &    1.68    &    83.79    &    -3.60    \\
15    &    18.01    &    29.77    &    1.25    &    1.290    &    2.594    &    1.492 &    31.63    &    68.37    &    4.05    &    2.51    &    84.72    &    -3.94    \\
\hline 
\multicolumn{13}{c}{a$_{ini}=1.5$ [au], $P\simeq1.3$ [yr]}\\
\hline 
1    &    1.47    &    2.42    &    0.17    &    1.298    &    3.242    &    0.547 &    58.25    &    41.75    &    0.65    &    0.11    &    61.45    &    -1.54    \\
5    &    7.33    &    12.12    &    0.33    &    1.296    &    3.058    &    0.891 &    43.07    &    56.93    &    2.07    &    0.56    &    72.58    &    -2.02    \\
10    &    14.69    &    24.28    &    0.56    &    1.295    &    2.959    &    1.014 &    37.73    &    62.27    &    3.33    &    1.12    &    76.62    &    -2.33    \\
15    &    22.06    &    36.46    &    0.81    &    1.295    &    2.894    &    1.014 &    34.48    &    65.52    &    4.36    &    1.67    &    78.74    &    -2.55    \\
\hline 
\hline 
\multicolumn{13}{c}{ }\\
\multicolumn{13}{c}{2.5 $M_{\odot}$, $\Omega_{ini}$/ $\Omega_{crit}$=0.1, $\mathcal{L}_{0,\star} = 0.92$  [$10^{50}$ g cm$^2$ s$^{-1}$], Z=0.02}\\
\multicolumn{13}{c}{ }\\
\hline 
\multicolumn{13}{c}{a$_{ini}=0.5$ [au], $P\simeq0.22$ [yr]}\\
\hline 
1    &    0.95    &    1.03    &    -        &    -        &    -        & - &    -        &    4.30    &    0.71    &    -    &    -    &    -    \\
5    &    4.73    &    5.14    &    0.19    &    0.642    &    2.415    &    1.875 &    45.09    &    54.91    &    0.39    &    2.10    &    89.72    &    -8.15    \\
10    &    9.48    &    10.29    &    0.20    &    0.642    &    2.355    &    1.875 &    39.37    &    60.63    &    0.61    &    4.19    &    89.78    &    -8.67    \\
15    &    14.23    &    15.45    &    0.20    &    0.641    &    2.319    &    2.004 &    36.18    &    63.82    &    0.79    &    6.28    &    89.83    &    -9.17    \\
\hline 
\multicolumn{13}{c}{a$_{ini}=1$ [au], $P\simeq0.63$ [yr]}\\
\hline 
1    &    1.34    &    1.45    &    -    &    -    &    -                &    -    &    -   &    0.46    &    0.23    &    -        &    -        &    -        \\
5    &    6.69    &    7.27    &    -    &    -    &    -                &    -    &    -   &    2.04    &    1.00    &    -        &    -        &    -        \\
10    &    13.40    &    14.55    &    -    &    -    &    -                &    -    &    -   &    4.14    &    2.03    &    -        &    -        &    -        \\
15    &    20.12    &    21.85    &    -    &    -    &    -                &    -    &    -   &    6.35    &    3.10    &    -        &    -        &    -        \\
\hline 
\multicolumn{13}{c}{a$_{ini}=1.5$ [au], $P\simeq1.16$ [yr]}\\
\hline 
1    &    1.64    &    1.78    &    -    &    -    &    -                &    -    &    -   &    0.24    &    0.12    &    -        &    -        &    -        \\
5    &    8.20    &    8.90    &    -    &    -    &    -                &    -    &    -   &    1.08    &    0.54    &    -        &    -        &    -        \\
10    &    16.41    &    17.82    &    -    &    -    &    -                &    -    &    -   &    2.16    &    1.08    &    -        &    -        &    -        \\
15    &    24.64    &    26.76    &    -    &    -    &    -                &    -    &    -   &    3.27    &    1.64    &    -        &    -        &    -        \\
\hline\hline 
\end{tabular}
}}
\label{table:ap1}
\end{table*}

\begin{table*}
\scriptsize{
\caption{Same as Table \hyperref[table:ap1]{A.1} but for stars with an initial surface angular velocity equal to 50\% the critical angular velocity.} 
\resizebox{18cm}{!} {
\begin{tabular}{ccccccccccccc}
\hline\hline 
\multicolumn{13}{c}{1.5 $M_{\odot}$, $\Omega_{ini}$/ $\Omega_{crit}$=0.5, $\mathcal{L}_{0,\star} = 2.16$  [$10^{50}$ g cm$^2$ s$^{-1}$], Z=0.02}\\
\hline 
$M_{pl}$ & $\mathcal{L}_{0,pl}$  & $\mathcal{L}_{0,pl}/\mathcal{L}_{0,\star}$ &$t_{migr}$ & Age$_{eng}$& $\log L_{eng}/L_{\odot}$& $\log g_{eng} $ & $\mathcal{L}_{eng}$ &$\mathcal{L}_{migr}$   & $f_{notides}$ & $E_{kin}(Age_{eng})$ &  $\mathcal{L}_{env}(Age_{eng})$ & $E_{pot}(Age_{eng})$  \\ 
 $[M_{J}]$ & [$10^{50}$ g cm$^2$ s$^{-1}$] & &[Myr] & [Gyr]& & &[$\% \ \mathcal{L}_{0,pl}$]& [$\% \ \mathcal{L}_{0,pl}$]& [$\% \ \mathcal{L}_{migr}$]    & [$10^{44}$ erg] & [$\% \ \mathcal{L}_{0,\star}$]& [$10^{47}$ erg] \\
 (1)&(2)&(3)&(4)&(5)&(6)&(7)&(8)&(9)&(10)&(11)&(12)&(13) \\
\hline 
\multicolumn{13}{c}{ }\\
\multicolumn{13}{c}{1.5 $M_{\odot}$, $\Omega_{ini}$/ $\Omega_{crit}$=0.5, $\mathcal{L}_{0,\star} = 2.16$  [$10^{50}$ g cm$^2$ s$^{-1}$], Z=0.02}\\
\multicolumn{13}{c}{ }\\
\hline 
\multicolumn{13}{c}{a$_{ini}=0.5$ [au], $P\simeq0.29$ [yr]}\\
\hline 
1    &    0.73    &    0.34    &    0.77    &    3.852    &    2.318    &    1.639 &    41.56    &    58.44    &    0.95    &    0.25     &    73.90    &    -3.21    \\
5    &    3.67    &    1.70    &    2.17    &    3.844    &    2.102    &    1.917 &    30.81    &    69.19    &    3.07    &    1.26    &    75.18    &    -4.27    \\
10    &    7.35    &    3.40    &    5.77    &    3.837    &    1.976    &    2.093 &    26.22    &    73.78    &    5.12    &    2.51    &    75.71    &    -5.05    \\
15    &    11.04    &    5.11    &    13.85    &    3.832    &    1.885    &    2.192 &    23.60    &    76.40    &    6.82    &    3.75    &    76.01    &    -5.70    \\
\hline 
\multicolumn{13}{c}{a$_{ini}=1$ [au], $P\simeq0.82$ [yr]}\\
\hline 
1    &    1.04    &    0.48    &    0.28    &    3.862    &    2.859    &    0.933 &    51.03    &    48.97    &    0.89    &    0.13    &    65.80    &    -1.48    \\
5    &    5.19    &    2.40    &    0.72    &    3.859    &    2.661    &    1.217 &    37.80    &    62.20    &    2.95    &    0.63    &    70.06    &    -1.98    \\
10    &    10.39    &    4.81    &    1.89    &    3.857    &    2.545    &    1.375 &    33.83    &    66.17    &    4.91    &    1.26    &    71.70    &    -2.37    \\
15    &    15.62    &    7.23    &    3.94    &    3.855    &    2.458    &    1.491 &    29.22    &    70.78    &    6.47    &    1.88    &    72.67    &    -2.68    \\
\hline 
\multicolumn{13}{c}{a$_{ini}=1.5$ [au], $P\simeq1.5$ [yr]}\\
\hline 
1    &    1.27    &    0.59    &    0.20    &    3.865    &    3.168    &    0.529 &    54.11    &    45.89    &    0.94    &    0.08    &    53.12    &    -0.95    \\
5    &    6.35    &    2.94    &    0.48    &    3.863    &    2.959    &    0.815 &    40.50    &    59.50    &    2.99    &    0.42    &    62.65    &    -1.28    \\
10    &    12.73    &    5.89    &    1.07    &    3.862    &    2.848    &    0.933 &    38.51    &    61.49    &    4.86    &    0.84    &    66.07    &    -1.50    \\
15    &    19.12    &    8.85    &    2.03    &    3.861    &    2.770    &    1.042 &    31.59    &    68.41    &    6.36    &    1.25    &    68.01    &    -1.71    \\
\hline 
\multicolumn{13}{c}{ }\\
\multicolumn{13}{c}{1.7 $M_{\odot}$, $\Omega_{ini}$/ $\Omega_{crit}$=0.5, $\mathcal{L}_{0,\star} = 2.39$  [$10^{50}$ g cm$^2$ s$^{-1}$], Z=0.02}\\
\multicolumn{13}{c}{ }\\
\hline 
\multicolumn{13}{c}{a$_{ini}=0.5$ [au], $P\simeq0.27$ [yr]}\\
\hline 
1    &    0.78    &    0.33    &    0.63    &    2.564    &    2.369    &    1.663 &    44.15      &    55.85    &    0.73    &    0.29    &    79.97    &    -3.88    \\
5    &    3.91    &    1.63    &    1.41    &    2.558    &    2.168    &    1.914 &    32.5       &    67.50    &    2.36    &    1.43    &    80.95    &    -5.08    \\
10    &    7.82    &    3.27    &    3.11    &    2.553    &    2.056    &    2.078 &    28.22      &    71.78    &    3.95    &    2.85    &    81.32    &    -5.91    \\
15    &    11.75    &    4.92    &    6.93    &    2.549    &    1.969    &    2.187 &    28.56      &    71.44    &    5.25    &    4.26    &    81.54    &    -6.66    \\
\hline 
\multicolumn{13}{c}{a$_{ini}=1$ [au], $P\simeq0.77$ [yr]}\\
\hline 
1    &    1.10    &    0.46    &    0.26    &    2.573    &    2.904    &    0.964 &     55.26     &    44.74    &    0.73    &    0.14    &    73.33    &    -1.83    \\
5    &    5.52    &    2.31    &    0.54    &    2.571    &    2.718    &    1.177 &    40.27      &    59.73    &    2.41    &    0.71    &    76.74    &    -2.39    \\
10    &    11.06    &    4.63    &    1.14    &    2.569    &    2.613    &    1.359 &    34.89      &    65.11    &    4.00    &    1.42    &    78.02    &    -2.82    \\
15    &    16.61    &    6.95    &    2.06    &    2.568    &    2.536    &    1.432 &    33.11      &    66.89    &    5.29    &    2.13    &    78.76    &    -3.12    \\
\hline 
\multicolumn{13}{c}{a$_{ini}=1.5$ [au], $P\simeq1.41$ [yr]}\\
\hline 
1    &    1.35    &    0.57    &    0.17    &    2.576    &    3.211    &    0.510 &     54.67     &    45.33    &    0.79    &    0.10    &    62.67    &    -1.17    \\
5    &    6.76    &    2.83    &    0.37    &    2.574    &    3.013    &    0.835 &     43.33     &    56.67    &    2.52    &    0.48    &    70.47    &    -1.54    \\
10    &    13.55    &    5.67    &    0.73    &    2.573    &    2.910    &    0.964 &     37.01     &    62.99    &    4.10    &    0.95    &    73.19    &    -1.82    \\
15    &    20.35    &    8.51    &    1.22    &    2.572    &    2.838    &    1.075 &    33.53      &    66.47    &    5.37    &    1.42    &    74.73    &    -2.01    \\
\hline 
\multicolumn{13}{c}{ }\\
\multicolumn{13}{c}{2.0 $M_{\odot}$, $\Omega_{ini}$/ $\Omega_{crit}$=0.5, $\mathcal{L}_{0,\star} = 3.06$  [$10^{50}$ g cm$^2$ s$^{-1}$], Z=0.02}\\
\multicolumn{13}{c}{ }\\
\hline 
\multicolumn{13}{c}{a$_{ini}=0.5$ [au], $P\simeq0.25$ [yr]}\\
\hline 
1    &    0.85    &    0.28    &    0.47    &    1.516    &    2.435    &    1.699 &    48.59    &    51.41    &    0.49    &    0.34    &    87.32    &    -4.97    \\
5    &    4.23    &    1.38    &    1.16    &    1.511    &    2.232    &    1.968 &    33.66    &    66.34    &    1.53    &    1.68    &    88.01    &    -6.51    \\
10    &    8.48    &    2.77    &    6.15    &    1.505    &    2.034    &    2.230 &    26.37    &    73.63    &    2.25    &    3.351    &    88.40    &    -8.62    \\
15    &    12.73    &    4.16    &    0.78    &    1.498    &    2.144    &    2.077 &    27.07    &    72.93    &    2.01    &    5.024    &    88.70    &    -7.44    \\
\hline 
\multicolumn{13}{c}{a$_{ini}=1$ [au], $P\simeq0.71$ [yr]}\\
\hline 
1    &    1.20    &    0.39    &    0.21    &    1.524    &    2.963    &    0.954 &    54.05    &    45.95    &    0.55    &    0.17    &    82.30    &    -2.36    \\
5    &    5.99    &    1.96    &    0.38    &    1.522    &    2.793    &    1.208 &    40.59    &    59.41    &    1.80    &    0.84    &    84.76    &    -3.06    \\
10    &    11.99    &    3.92    &    0.62    &    1.521    &    2.699    &    1.316 &    36.51    &    63.49    &    2.95    &    1.68    &    85.71    &    -3.51    \\
15    &    18.01    &    5.89    &    0.93    &    1.520    &    2.635    &    1.403 &    32.14    &    67.86    &    3.90    &    2.519    &    86.21    &    -3.80    \\
\hline 
\multicolumn{13}{c}{a$_{ini}=1.5$ [au], $P\simeq1.3$ [yr]}\\
\hline 
1    &    1.47    &    0.48    &    -        &    -        &    -        & -    &    -        &    4.49    &    0.50        &    -   &    -    &    -    \\
5    &    7.33    &    2.40    &    0.28    &    1.525    &    3.083    &    0.795 &    44.57    &    50.10    &    1.98    &    0.56    &    79.64    &    -1.95    \\
10    &    14.69    &    4.80    &    0.45    &    1.524    &    2.988    &    0.954 &    40.75    &    61.70    &    3.21    &    1.12    &    81.99    &    -2.31    \\
15    &    22.06    &    7.21    &    0.65    &    1.523    &    2.926    &    1.086 &    37.42    &    62.91    &    4.21    &    1.67    &    83.57    &    -2.65    \\
\hline 
\multicolumn{13}{c}{ }\\
\multicolumn{13}{c}{2.5 $M_{\odot}$, $\Omega_{ini}$/ $\Omega_{crit}$=0.5, $\mathcal{L}_{0,\star} = 4.65$  [$10^{50}$ g cm$^2$ s$^{-1}$], Z=0.02}\\
\multicolumn{13}{c}{ }\\
\hline 
\multicolumn{13}{c}{a$_{ini}=0.5$ [au], $P\simeq0.22$ [yr]}\\
\hline 
1    &    0.95    &    0.20    &    -        &    -        &    -        & -    &    -        &    3.35    &    0.77    &    -        &    -        &    -        \\
5    &    4.73    &    1.02    &    0.15    &    0.761    &    2.456    &    1.880 &    51.33    &    48.67    &    0.37    &    2.10    &    86.59    &    -10.4    \\
10    &    9.48    &    2.04    &    0.14    &    0.761    &    2.392    &    1.880 &    40.08    &    59.92    &    0.59    &    4.19    &    86.59    &    -10.4    \\
15    &    14.23    &    3.06    &    0.15    &    0.761    &    2.359    &    1.880 &    39.25    &    60.75    &    0.76    &    6.28    &    86.59    &    -10.4    \\
\hline 
\multicolumn{13}{c}{a$_{ini}=1$ [au], $P\simeq0.63$ [yr]}\\
\hline 
1    &    1.34    &    0.29    &    -    &    -    &    -                &    -    &    -        &    0.50    &    0.25    &    -        &    -        &    -        \\
5    &    6.69    &    1.44    &    -    &    -    &    -                &    -    &    -        &    2.25    &    1.12    &    -        &    -        &    -        \\
10    &    13.40    &    2.88    &    -    &    -    &    -                &    -    &    -        &    4.58    &    2.28    &    -        &    -        &    -        \\
15    &    20.12    &    4.33    &    -    &    -    &    -                &    -    &    -        &    7.03    &    3.50    &    -        &    -        &    -        \\
\hline 
\multicolumn{13}{c}{a$_{ini}=1.5$ [au], $P\simeq1.16$ [yr]}\\
\hline 
1    &    1.64    &    0.35    &    -    &    -    &    -                &    -    &    -        &    0.27    &    0.13    &    -        &    -        &    -        \\
5    &    8.20    &    1.76    &    -    &    -    &    -                &    -    &    -        &    1.20    &    0.60    &    -        &    -        &    -        \\
10    &    16.41    &    3.53    &    -    &    -    &    -                &    -    &    -        &    2.42    &    1.21    &    -        &    -        &    -        \\
15    &    24.64    &    5.30    &    -    &    -    &    -                &    -    &    -        &    3.67    &    1.84    &    -        &    -        &    -        \\
\hline\hline 
\end{tabular}
}}
\label{table:ap2}
\end{table*}

\end{appendix}

\bibliographystyle{aa} 
\bibliography{biblio} 

\begin{thebibliography}{53}
\expandafter\ifx\csname natexlab\endcsname\relax\def\natexlab#1{#1}\fi

\bibitem[{{Adam{\'o}w} {et~al.}(2012){Adam{\'o}w}, {Niedzielski}, {Villaver},
  {Nowak}, \& {Wolszczan}}]{adamow12}
{Adam{\'o}w}, M., {Niedzielski}, A., {Villaver}, E., {Nowak}, G., \&
  {Wolszczan}, A. 2012, \apjl, 754, L15

\bibitem[{{Akeson} {et~al.}(2013){Akeson}, {Chen}, {Ciardi}, {Crane}, {Good},
  {Harbut}, {Jackson}, {Kane}, {Laity}, {Leifer}, {Lynn}, {McElroy}, {Papin},
  {Plavchan}, {Ram{\'{\i}}rez}, {Rey}, {von Braun}, {Wittman}, {Abajian},
  {Ali}, {Beichman}, {Beekley}, {Berriman}, {Berukoff}, {Bryden}, {Chan},
  {Groom}, {Lau}, {Payne}, {Regelson}, {Saucedo}, {Schmitz}, {Stauffer},
  {Wyatt}, \& {Zhang}}]{akeson13}
{Akeson}, R.~L., {Chen}, X., {Ciardi}, D., {et~al.} 2013, \pasp, 125, 989

\bibitem[{{Alexander} {et~al.}(1976){Alexander}, {Chau}, \&
  {Henriksen}}]{alexander76}
{Alexander}, M.~E., {Chau}, W.~Y., \& {Henriksen}, R.~N. 1976, \apj, 204, 879

\bibitem[{{Bear} \& {Soker}(2011)}]{bear11}
{Bear}, E. \& {Soker}, N. 2011, \mnras, 414, 1788

\bibitem[{{Bouvier}(2008)}]{Bouvier2008}
{Bouvier}, J. 2008, \aap, 489, L53

\bibitem[{{Carlberg}(2014)}]{carlberg14}
{Carlberg}, J.~K. 2014, \aj, 147, 138

\bibitem[{{Carlberg} {et~al.}(2012){Carlberg}, {Cunha}, {Smith}, \&
  {Majewski}}]{carlberg12}
{Carlberg}, J.~K., {Cunha}, K., {Smith}, V.~V., \& {Majewski}, S.~R. 2012,
  \apj, 757, 109

\bibitem[{{Carlberg} {et~al.}(2009){Carlberg}, {Majewski}, \&
  {Arras}}]{carlberg09}
{Carlberg}, J.~K., {Majewski}, S.~R., \& {Arras}, P. 2009, \apj, 700, 832

\bibitem[{{Carlberg} {et~al.}(2011){Carlberg}, {Majewski}, {Patterson},
  {Bizyaev}, {Smith}, \& {Cunha}}]{carlberg11}
{Carlberg}, J.~K., {Majewski}, S.~R., {Patterson}, R.~J., {et~al.} 2011, \apj,
  732, 39

\bibitem[{{Carlberg} {et~al.}(2010){Carlberg}, {Smith}, {Cunha}, {Majewski}, \&
  {Rood}}]{carlberg10}
{Carlberg}, J.~K., {Smith}, V.~V., {Cunha}, K., {Majewski}, S.~R., \& {Rood},
  R.~T. 2010, \apjl, 723, L103

\bibitem[{{Chapman} \& {Ferraro}(1930)}]{chapman30}
{Chapman}, S. \& {Ferraro}, V.~C.~A. 1930, \nat, 126, 129

\bibitem[{{Donati} \& {Landstreet}(2009)}]{Donati2009}
{Donati}, J.-F. \& {Landstreet}, J.~D. 2009, \araa, 47, 333

\bibitem[{{Eggenberger} {et~al.}(2008){Eggenberger}, {Meynet}, {Maeder},
  {Hirschi}, {Charbonnel}, {Talon}, \& {Ekstr{\"o}m}}]{egg08}
{Eggenberger}, P., {Meynet}, G., {Maeder}, A., {et~al.} 2008, \apss, 316, 43

\bibitem[{{Eggenberger} {et~al.}(2010){Eggenberger}, {Miglio}, {Montalban},
  {Moreira}, {Noels}, {Meynet}, \& {Maeder}}]{eggen10}
{Eggenberger}, P., {Miglio}, A., {Montalban}, J., {et~al.} 2010, \aap, 509, A72

\bibitem[{{Ekstr{\"o}m} {et~al.}(2012){Ekstr{\"o}m}, {Georgy}, {Eggenberger},
  {Meynet}, {Mowlavi}, {Wyttenbach}, {Granada}, {Decressin}, {Hirschi},
  {Frischknecht}, {Charbonnel}, \& {Maeder}}]{ekstrom12}
{Ekstr{\"o}m}, S., {Georgy}, C., {Eggenberger}, P., {et~al.} 2012, \aap, 537,
  A146

\bibitem[{{Fekel} \& {Balachandran}(1993)}]{fekel93}
{Fekel}, F.~C. \& {Balachandran}, S. 1993, \apj, 403, 708

\bibitem[{{Han} {et~al.}(2014){Han}, {Wang}, {Wright}, {Feng}, {Zhao},
  {Fakhouri}, {Brown}, \& {Hancock}}]{Han2014}
{Han}, E., {Wang}, S.~X., {Wright}, J.~T., {et~al.} 2014, \pasp, 126, 827

\bibitem[{{Kunitomo} {et~al.}(2011){Kunitomo}, {Ikoma}, {Sato}, {Katsuta}, \&
  {Ida}}]{kunimoto11}
{Kunitomo}, M., {Ikoma}, M., {Sato}, B., {Katsuta}, Y., \& {Ida}, S. 2011,
  \apj, 737, 66

\bibitem[{Lang(2011)}]{cambridge11}
Lang, K. 2011, The Cambridge Guide to the Solar System (Cambridge University
  Press)

\bibitem[{{Livio} \& {Soker}(1984{\natexlab{a}})}]{livio84}
{Livio}, M. \& {Soker}, N. 1984{\natexlab{a}}, \mnras, 208, 783

\bibitem[{{Livio} \& {Soker}(1984{\natexlab{b}})}]{livio_soker84}
{Livio}, M. \& {Soker}, N. 1984{\natexlab{b}}, \mnras, 208, 763

\bibitem[{{Maeder} \& {Meynet}(1989{\natexlab{a}})}]{maeder89}
{Maeder}, A. \& {Meynet}, G. 1989{\natexlab{a}}, \aap, 210, 155

\bibitem[{{Maeder} \& {Meynet}(1989{\natexlab{b}})}]{MM1989}
{Maeder}, A. \& {Meynet}, G. 1989{\natexlab{b}}, \aap, 210, 155

\bibitem[{{Maeder} \& {Meynet}(2012)}]{MM2012}
{Maeder}, A. \& {Meynet}, G. 2012, Reviews of Modern Physics, 84, 25

\bibitem[{{Maeder} \& {Zahn}(1998)}]{MZ1998}
{Maeder}, A. \& {Zahn}, J.-P. 1998, \aap, 334, 1000

\bibitem[{{Maldonado} \& {Villaver}(2016)}]{Maldonado2016}
{Maldonado}, J. \& {Villaver}, E. 2016, \aap, 588, A98

\bibitem[{{Maldonado} {et~al.}(2013){Maldonado}, {Villaver}, \&
  {Eiroa}}]{maldonado13}
{Maldonado}, J., {Villaver}, E., \& {Eiroa}, C. 2013, \aap, 554, A84

\bibitem[{{Massarotti} {et~al.}(2008){Massarotti}, {Latham}, {Stefanik}, \&
  {Fogel}}]{massarotti08}
{Massarotti}, A., {Latham}, D.~W., {Stefanik}, R.~P., \& {Fogel}, J. 2008, \aj,
  135, 209

\bibitem[{{Mustill} \& {Villaver}(2012)}]{mustill12}
{Mustill}, A.~J. \& {Villaver}, E. 2012, \apj, 761, 121

\bibitem[{{Nordhaus} \& {Spiegel}(2013)}]{nordhaus13}
{Nordhaus}, J. \& {Spiegel}, D.~S. 2013, \mnras, 432, 500

\bibitem[{{Nordhaus} {et~al.}(2010){Nordhaus}, {Spiegel}, {Ibgui}, {Goodman},
  \& {Burrows}}]{nordhaus10}
{Nordhaus}, J., {Spiegel}, D.~S., {Ibgui}, L., {Goodman}, J., \& {Burrows}, A.
  2010, \mnras, 408, 631

\bibitem[{{Rasio} {et~al.}(1996){Rasio}, {Tout}, {Lubow}, \& {Livio}}]{rasio96}
{Rasio}, F.~A., {Tout}, C.~A., {Lubow}, S.~H., \& {Livio}, M. 1996, \apj, 470,
  1187

\bibitem[{{Reimers}(1975)}]{rei75}
{Reimers}, D. 1975, Memoires of the Societe Royale des Sciences de Liege, 8,
  369

\bibitem[{{Sackmann} {et~al.}(1993){Sackmann}, {Boothroyd}, \&
  {Kraemer}}]{sackmann93}
{Sackmann}, I.-J., {Boothroyd}, A.~I., \& {Kraemer}, K.~E. 1993, \apj, 418, 457

\bibitem[{{Santos} {et~al.}(2001){Santos}, {Israelian}, \&
  {Mayor}}]{santos2001}
{Santos}, N.~C., {Israelian}, G., \& {Mayor}, M. 2001, ArXiv Astrophysics
  e-prints

\bibitem[{{Santos} {et~al.}(2004){Santos}, {Israelian}, \&
  {Mayor}}]{santos2004}
{Santos}, N.~C., {Israelian}, G., \& {Mayor}, M. 2004, VizieR Online Data
  Catalog, 341, 51153

\bibitem[{{Sato} {et~al.}(2008){Sato}, {Toyota}, {Omiya}, {Izumiura}, {Kambe},
  {Masuda}, {Takeda}, {Itoh}, {Ando}, {Yoshida}, {Kokubo}, \& {Ida}}]{sato08}
{Sato}, B., {Toyota}, E., {Omiya}, M., {et~al.} 2008, \pasj, 60, 1317

\bibitem[{{Siess} \& {Livio}(1999{\natexlab{a}})}]{siess99I}
{Siess}, L. \& {Livio}, M. 1999{\natexlab{a}}, \mnras, 304, 925

\bibitem[{{Siess} \& {Livio}(1999{\natexlab{b}})}]{siess99II}
{Siess}, L. \& {Livio}, M. 1999{\natexlab{b}}, \mnras, 308, 1133

\bibitem[{{Simon} \& {Drake}(1989)}]{simon89}
{Simon}, T. \& {Drake}, S.~A. 1989, \apj, 346, 303

\bibitem[{{Soker} {et~al.}(1984){Soker}, {Livio}, \& {Harpaz}}]{soker84}
{Soker}, N., {Livio}, M., \& {Harpaz}, A. 1984, \mnras, 210, 189

\bibitem[{{Sousa} {et~al.}(2011){Sousa}, {Santos}, {Israelian}, {Mayor}, \&
  {Udry}}]{sousa2011}
{Sousa}, S.~G., {Santos}, N.~C., {Israelian}, G., {Mayor}, M., \& {Udry}, S.
  2011, \aap, 533, A141

\bibitem[{{Talon}(1997)}]{tal97}
{Talon}, S. 1997, PhD thesis, , Observatoire de Paris, (1997), 187 pages

\bibitem[{{Vidotto} {et~al.}(2014){Vidotto}, {Jardine}, {Morin}, {Donati},
  {Opher}, \& {Gombosi}}]{Vidotto2014}
{Vidotto}, A.~A., {Jardine}, M., {Morin}, J., {et~al.} 2014, \mnras, 438, 1162

\bibitem[{{Villaver} \& {Livio}(2007)}]{villaver07}
{Villaver}, E. \& {Livio}, M. 2007, \apj, 661, 1192

\bibitem[{{Villaver} \& {Livio}(2009)}]{villaver09}
{Villaver}, E. \& {Livio}, M. 2009, \apjl, 705, L81

\bibitem[{{Villaver} {et~al.}(2014){Villaver}, {Livio}, {Mustill}, \&
  {Siess}}]{villaver14}
{Villaver}, E., {Livio}, M., {Mustill}, A.~J., \& {Siess}, L. 2014, \apj, 794,
  3

\bibitem[{{Zahn}(1966)}]{zahn66}
{Zahn}, J.~P. 1966, Annales d'Astrophysique, 29, 489

\bibitem[{{Zahn}(1977)}]{zahn77}
{Zahn}, J.-P. 1977, \aap, 57, 383

\bibitem[{{Zahn}(1989)}]{zahn89}
{Zahn}, J.-P. 1989, \aap, 220, 112

\bibitem[{{Zahn}(1992)}]{zahn92}
{Zahn}, J.-P. 1992, \aap, 265, 115

\bibitem[{{Zapolsky} \& {Salpeter}(1969)}]{zapolsky69}
{Zapolsky}, H.~S. \& {Salpeter}, E.~E. 1969, \apj, 158, 809

\bibitem[{{Zorec} \& {Royer}(2012)}]{ZR2012}
{Zorec}, J. \& {Royer}, F. 2012, \aap, 537, A120

\end{thebibliography}

\end{document}